\documentclass[cernpreprint, atlasdraft=false, texlive=2023, UKenglish, texmf, orcidlogo]{atlasdoc}
 
\usepackage{atlaspackage}
\usepackage{atlasbiblatex}
 
\usepackage{multirow}	
 
\usepackage{caption}
\usepackage{subcaption}
 
\usepackage{atlasphysics}

\graphicspath{{logos/}{figures/}}
 
\usepackage{ANA-EXOT-2018-54-PAPER-defs}

\AtlasTitle{Search for heavy long-lived multi-charged particles in the full LHC Run~2 $pp$ collision data at $\sqrt{s} = 13$~TeV using the ATLAS detector}

\AtlasAbstract{
A search for heavy long-lived multi-charged particles is performed using the ATLAS detector at the LHC\@.
Data collected in 2015--2018 at $\sqrt{s} = 13$~TeV from $pp$ collisions corresponding to an integrated luminosity
of 139~fb$^{-1}$ are examined.
Particles producing anomalously high ionization, consistent with long-lived spin-\textonehalf{} massive particles
with electric charges from ${|q|=2e}$ to ${|q|=7e}$ are searched for.
No statistically significant evidence of such particles is observed,
and \SI{95}{\percent} confidence level cross-section upper limits are calculated and interpreted as
the lower mass limits for a Drell--Yan plus photon-fusion production mode.
The least stringent limit, $1060$~GeV, is obtained for ${|q|=2e}$ particles, and the most stringent one,
$1600$~GeV, is for ${|q|=6e}$ particles.
}

\AtlasRefCode{EXOT-2018-54}
 
\PreprintIdNumber{CERN-EP-2023-017}

\AtlasJournalRef{Phys. Lett. B 847 (2023) 138316}
\AtlasDOI{10.1016/j.physletb.2023.138316}

\hypersetup{pdftitle={ATLAS document},pdfauthor={The ATLAS Collaboration}}
 
\begin{document}
 
\maketitle

\section{Introduction}
 
This Letter describes a search for heavy long-lived multi-charged particles (MCPs)
in ${\rts=\SI{13}{\TeV{}}}$ proton--proton collision data collected in 2015--2018 by the ATLAS detector
at the CERN Large Hadron Collider (LHC)~\cite{Evans:2008zzb}.
The search, conducted on a sample of data corresponding to an integrated luminosity of \SI{139}{\ifb},
is performed in the MCP mass range from $500$ to \SI{2000}{\GeV{}},
for electric charges\footnote{Wherever a charge is quoted for exotic particles, the charge
conjugate state is also implied.} ${|q|=ze}$, with integer charge numbers ${2 \leq z \leq 7}$.
The mass range starts from the mass value excluded in the previous iterations of this search~\cite{EXOT-2013-20,EXOT-2017-13}
and ends with a value driven by the expected sensitivity of the current search.
An observation of such particles, possessing an electric charge above the elementary
charge $e$, would be a signature for physics beyond the Standard Model (SM).
 
Several theoretical models predict multi-charged particles.
AC-leptons (anion-like and cation-like exotic ions of unknown matter, respectively)~\cite{Fargion:2005ep},
as predicted by the almost-commutative model~\cite{Stephan:2005uj},
are pairs of SU(2) electroweak singlets with opposite electromagnetic charges and no
other SM gauge charges, which makes them behave as heavy stable charged leptons.
Technibaryons, predicted by the walking-technicolor model~\cite{Sannino:2004qp}, are
Goldstone bosons made of two techniquarks or two anti-techniquarks with
an arbitrary value of the electric charge.
The lightest technibaryon is expected to be stable in
the absence of processes violating the technibaryon number conservation law.
Doubly charged Higgs bosons are predicted by the left--right symmetric model~\cite{Mohapatra:1974hk}
in Higgs boson triplets in a model postulating a right-handed version of the weak interaction.
Its gauge symmetry would be spontaneously broken at a high mass scale, leading to the observed parity violation
in the weak-interaction sector of the SM\@.
Only leptonic decay modes would be characteristic of such particles~\cite{Aulakh:1997ba}.
The ${H^{\pm\pm} \rightarrow W^{\pm}W^{\pm}}$ decays are assumed to be suppressed.
The supersymmetric left--right model~\cite{Aulakh:1997ba} predicts a long-lived light $H^{\pm\pm}$ boson
due to the lepton number conservation.
 
Any observation of the particles predicted by the first two models could have
implications for the formation of composite dark matter:
the doubly charged particles (or, more generally, particles with an even charge ${|q| = 2ne}$)
could explain some excesses (e.g., positron excess) observed in direct and indirect searches
for dark matter~\cite{PhysRevLett.110.141102,Khlopov:2011tn,Belotsky:2015gsa}.
So far, no such particles have been observed in cosmic-ray~\cite{Cecchini:2008su} or collider searches,
including several recent searches at the Tevatron~\cite{Acosta:2005np} and
the LHC~\cite{EXOT-2013-20,EXOT-2017-13,CMS-EXO-12-026,CMS-EXO-15-010}.
 
In this search, the MCPs are assumed to live long enough to traverse the entire ATLAS detector without decaying,
and thus the analysis exploits their muon-like signature, making a muon trigger a natural choice
(although other triggers are also used).
Given their high mass compared to a muon and their high ionization losses, \dEdx{}, compared to a ${z=1}$ particle,
they may not be triggered in the correct bunch crossing by a conventional muon trigger, as discussed in \Sect{\ref{sec:sigEff}}.
The addition of a missing-transverse-momentum trigger and of a `late-muon' trigger partially mitigates this issue.
The offline analysis searches for high-\pT{} muon-like tracks with high \dEdx{} values in several subdetector systems.
High \dEdx{} values arise from both higher electric charges and lower velocities of such particles
compared to most of the SM particles produced at the LHC.
The background expected from SM processes (largely high-\pT{} muons from decays of $Z$ and $W$~bosons, $\tau$-leptons, etc.)
is estimated by using a data-driven technique.
The results are interpreted in terms of upper limits set on the production cross-section as a function of mass
for different charge hypotheses, and in terms of lower mass limits as a function of charge.
These results supersede those of a previous search using a smaller \SI{13}{\TeV{}} data sample~\cite{EXOT-2017-13}.
Apart from a data sample four times larger, the main changes are related
to improvements in the production model (the addition of the photon-fusion production mode and the
virtual $Z^0$ exchange for the Drell--Yan mode~\cite{Song_2022}) and to the addition of the `late-muon' trigger.
 
This search complements recent ATLAS searches for heavy ${z=1}$ particles identifiable by their
high transverse momenta and anomalously large ionization losses in
the innermost tracking detector~\cite{SUSY-2016-31,ATLAS:2022pib}.
 
\section{ATLAS detector}
 
The ATLAS detector~\cite{PERF-2007-01} covers nearly the entire solid angle around the collision point.\footnote{ATLAS
uses a right-handed coordinate system with its origin at the nominal interaction point (IP) in the centre of the detector and
the $z$-axis along the beam pipe.
The $x$-axis points from the IP to the centre of the LHC ring, and the $y$-axis points upward.
Cylindrical coordinates ${(r,\phi)}$ are used in the transverse plane,
$\phi$ being the azimuthal angle around the $z$-axis.
The pseudorapidity is defined in terms of the polar angle $\theta$ as ${\eta=-\ln\tan(\theta/2)}$.
Angular distance is measured in units of $\Delta R \equiv \sqrt{(\Delta\eta)^{2} + (\Delta\phi)^{2}}$.}
The inner tracking detector (ID) consists of a silicon pixel detector (pixel), a silicon
microstrip detector (SCT) and a transition radiation tracker (TRT).
The pixel detector provides at least four precise space-point measurements per track.
At normal incidence, the average charge released by a minimum-ionizing particle (MIP) in a pixel sensor is typically
$\num{20000}~e^{-}$ ($\num{16000}~e^{-}$ for the IBL -- the insertable B-layer~\cite{PIX-2018-001,ATLAS-TDR-19},
an innermost layer with a different charge measurement) and the charge threshold is set to approximately
$\num{3500}~e^{-}$ ($\num{2500}~e^{-}$ for the IBL)~\cite{SUSY-2016-03}.
Only pixels where a signal exceeds the threshold are considered for further processing.
The dynamic range of the electronics measuring the ionization charge is typically limited to $\num{200000}~e^{-}$
($\num{30000}~e^{-}$ for the IBL).
Pixel electronic circuits shape the signals in such a way that the time above the threshold represents the amplitude
of the signal, which is proportional to the energy deposition of a charged particle.
In the IBL, an overflow bit is set if the time interval with the signal above the threshold exceeds its dynamic range,
in contrast to the other three layers of the pixel detector, where such a hit is lost due to limitations of the electronics
and no overflow bit is available.
However, since the charge released by a particle crossing the pixel detector is
rarely contained within just one pixel, the neighbouring pixels preserve the spatial information of this hit.
The SCT consists of four double-layer silicon sensors with binary readout architecture,
each with a small stereo angle, typically providing eight measurements per track.
The TRT, covering the pseudorapidity range ${\abseta<2.0}$, is a straw-tube
tracking detector capable of particle identification via transition-radiation and
ionization-energy-loss measurements~\cite{ATLAS-CONF-2011-128}.
A particle typically crosses $32$~straws.
The TRT straw signal processing uses low (LT) and high threshold (HT) discriminators.
LT discrimination defines a hit arrival time used for a space-point reconstruction for tracking.
The HT is designed to discriminate between energy depositions from transition-radiation photons and the energy loss of MIPs.
As \dEdx{} scales as $z^2$, MCPs would produce a large number of HT hits along their trajectories.
 
The ID is surrounded by a thin superconducting solenoid providing a \SI{2}{\tesla} axial magnetic field,
and by a high-granularity lead/liquid-argon (LAr) sampling electromagnetic calorimeter.
An iron/scintillator tile calorimeter provides hadronic-energy measurements in the central pseudorapidity region.
The endcap and forward regions are instrumented with LAr electromagnetic and hadronic calorimeters.
The calorimeter system is surrounded by a muon spectrometer (MS) incorporating three superconducting toroidal
magnet assemblies.
The MS is instrumented with tracking detectors designed to measure the momenta of muons.
Resistive-plate chambers (RPC) in the barrel region (${\abseta < 1.05}$) and thin-gap chambers (TGC) in the endcap
regions (${1.05 < \abseta < 2.4}$) provide signals for the trigger.
Monitored-drift-tube (MDT) chambers typically provide $20$--$25$ hits per crossing track in the pseudorapidity
range ${{\abseta < 2.7}}$, from which a high-precision momentum measurement is derived.
Each MDT readout channel, apart from time measurement, provides information about charge measured
in the first \SI{18.5}{\ns} following the initial threshold crossing~\cite{Arai:2008zzb}.
Cathode-strip chambers complement the tracking capabilities of the MDTs in the high-rate forward regions.
 
The amount of material in the ID varies from one-half to two radiation lengths.
The overall amount of material traversed by an MCP up to the last measurement surface,
which includes the calorimeters and the MS, may be as high as $75$ radiation lengths.
Muons typically lose \SI{3}{\GeV{}} penetrating the calorimeter system.
The energy loss for MCPs with charge $z$ would be $z^{2}$ times this value, i.e., up to \SI{150}{\GeV{}} for ${z=7}$.
 
The muon transverse momentum measured by the MS after the energy loss in the calorimeters is denoted by ${\pT^{\mu} / z}$,
while transverse momentum of charged particles measured by the combination of the ID and MS is denoted by ${\pT{} / z}$.
Charged-particle trajectories are reconstructed using standard algorithms~\cite{PERF-2015-08}.
Since these algorithms assume particles with unit electric charge,
the momenta of MCPs are underestimated by a factor $z$, as the track curvature is proportional to ${\pT{} / z}$.
 
A two-level trigger system is used to select interesting events~\cite{TRIG-2019-04}.
The first-level trigger (L1) is implemented in hardware and uses a subset of the detector information
to reduce the event rate to a design value of at most \SI{100}{\kHz}.
This is followed by the software-based high-level trigger, which further reduces the event rate to about \SI{1}{\kHz}.
 
An extensive software suite~\cite{ATL-SOFT-PUB-2021-001} is used in data simulation, in the
reconstruction and analysis of real and simulated data, in detector operations,
and in the trigger and data acquisition systems of the experiment.
 
\section{Samples of simulated events}
 
\begin{figure}[htb]
\begin{center}
\begin{subfigure}[b]{0.49\textwidth}
\centering
\includegraphics[width=\textwidth]{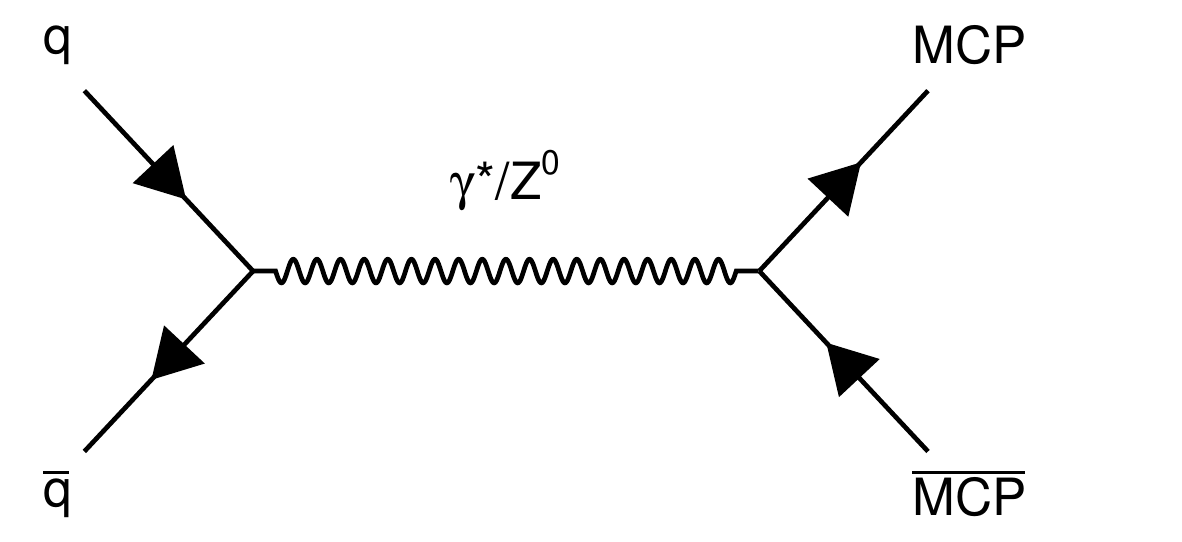}
\caption{}
\label{fig:feynmanDiagrams_DY}
\end{subfigure}
\hfill
\begin{subfigure}[b]{0.49\textwidth}
\centering
\includegraphics[width=\textwidth]{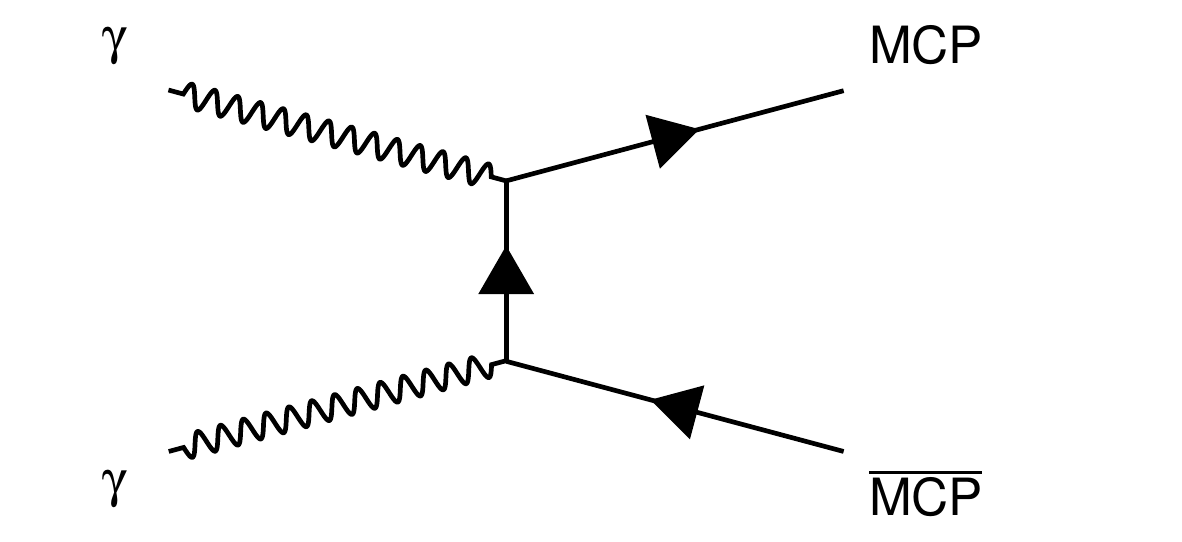}
\caption{}
\label{fig:feynmanDiagrams_PF}
\end{subfigure}
\caption{LO Feynman diagrams for MCP-generation processes: \subref{fig:feynmanDiagrams_DY} DY process and \subref{fig:feynmanDiagrams_PF} PF process ($t$-channel diagram).}
\label{fig:feynmanDiagrams}
\end{center}
\end{figure}

Benchmark samples of simulated events with MCPs were generated for a range of masses $m$
between $500$ and \SI{2000}{\GeV{}} in steps of \SI{300}{\GeV{}}, for charges $ze$ with ${z=2}$, $3$, \ldots, $7$.
Muon-like pairs of spin-\textonehalf{} MCPs were generated via two leading-order (LO) processes
implemented in \MGNLO[2.6.7]~\cite{Alwall:2014hca}:
the Drell--Yan (DY) process with both photon and $Z$~boson exchanges included, assuming the SM coupling of MCPs to the $Z$~boson
is proportional to the electric charge squared, and the photon-fusion (PF) process, as shown
in Feynman diagrams in \Fig{\ref{fig:feynmanDiagrams}}~\cite{Song_2022}.
 
This implementation of the production processes models the kinematic distributions and determines the cross-sections.
Cross-section values for spin-\textonehalf{} MCP pair production range from a few tenths of a picobarn
($m=\SI{500}{\GeV{}}$, ${z=7}$, PF mode)
down to a few tenths of an attobarn ($m=\SI{2000}{\GeV{}}$, ${z=2}$, PF mode).
Events were generated using the \NNPDF[2.3LO]~\cite{Carrazza:2013axa} and \LUXQED{}~\cite{Manohar:2017eqh}
parton distribution functions (PDFs), for DY and PF modes, respectively,
with the A14 set of tuned parameters~\cite{ATL-PHYS-PUB-2014-021}.
\PYTHIA[8.244]~\cite{Sjostrand:2014zea, Sjostrand:2006za} was used for hadronisation and underlying-event generation.
Samples with events produced via DY and PF modes (with the same mass and charge values) were merged
with production cross-section values acting as weights for each generated event.
PF-mode cross-sections dominate at high-charge values.
 
Samples of simulated events with muons from \Zmm{} decays were generated at the next-to-leading order (\NLO{})
using \POWHEGBOX{}~\cite{Nason:2004rx,Alioli:2008gx} interfaced to the \PYTHIA[8.186] parton shower model.
The \AZNLO{} tuned parameters~\cite{STDM-2012-23} were used,
with the \CTEQ[6L1] PDF set~\cite{Pumplin:2002vw} for the modelling of non-perturbative effects.
The \EVTGEN[1.2.0] program~\cite{Lange:2001uf} was used for the properties of $b$- and $c$-hadron decays.
 
A detailed \GEANT{} simulation~\cite{Agostinelli:2002hh, SOFT-2010-01} was used to model the response of the ATLAS detector.
Each simulated hard-scattering event was overlaid with simulated minimum-bias
events generated with \PYTHIA{} to emulate the data
distribution of multiple proton--proton collisions per bunch crossing (pile-up).
The simulated events are reconstructed and analyzed in the same way as the experimental data.
 
\section{Event and candidate selections}
 
The MCP identification relies on the ionization energy released by high-charge particles and measured
in the pixel, TRT, and MDT subdetector systems.
Acceptance is restricted to the pseudorapidity range of the TRT, ${\abseta{} < 2.0}$.
 
The selection is logically divided into four steps: trigger selection,
preselection, tight selection, and final selection.
While the first two steps rely on muon and missing-transverse-momentum signals as well as event topology, the tight and
final selection steps rely on the ionization estimators not available at the trigger level.
These estimators are introduced later in this section.
An event is considered to be a candidate event if it has at least one candidate
MCP (a reconstructed particle, which satisfies all the selection criteria).
 
\subsection{Trigger selection}
 
Events collected in 2015--2018 with a single-muon trigger with no isolation requirement and
a transverse-momentum threshold of ${\pT{}/z = \SI{50}{\GeV{}}}$ are considered.
This trigger is only sensitive to particles with velocity ${\beta=v/c>0.65}$ due to a timing window,
within which particles must reach the MS, which limits the trigger efficiency.
The efficiency of this trigger, averaged over all signal benchmark samples, is \SI{30}{\percent}.
It varies from \SI{6}{\percent} to \SI{54}{\percent} between the samples with $m=\SI{2000}{\GeV{}}$, $z=7$
and $m=\SI{500}{\GeV{}}$, $z=2$ MCPs, respectively.
 
To compensate for inefficiencies in the single-muon trigger, an additional calorimeter-based trigger that imposes
a threshold on the minimal magnitude of the missing transverse momentum, \MET{}, was employed.
The \MET{} threshold was set to \SI{70}{\GeV{}} in 2015 and was raised twice in 2016,
first to \SI{90}{\GeV{}} and later to \SI{110}{\GeV{}}.
Particles reconstructed in the MS are not accounted for in the trigger \MET{} calculation~\cite{TRIG-2019-01},
which only takes into account energy deposited in the calorimeters.
Large \MET{} originates from initial-state-radiation jets recoiling against the MCP pair.
The signal efficiency of the \MET{} trigger is \SIrange[range-phrase = --]{18}{25}{\percent}.
The combined efficiency of this trigger and of the single-muon trigger, averaged over all signal benchmark samples,
is \SI{39}{\percent}.
It varies from \SI{22}{\percent} to \SI{61}{\percent} between the samples with $m=\SI{2000}{\GeV{}}$, $z=7$
and $m=\SI{500}{\GeV{}}$, $z=2$ MCPs, respectively.

Finally, the third trigger used in the search is the `late-muon' trigger.
Its baseline algorithm is implemented within the L1 topological trigger~\cite{ATLAS:2021tnq}.
The traditional single-muon trigger cannot fire in any bunch crossing other than the current one,
and the magnitude of the \MET{} may not be high enough to fire the \MET{}-based trigger.
The late-muon trigger fires in events with a ${\pT{} > \SI{50}{\GeV{}}}$ jet in the current bunch-crossing
and a ${\pT{}/z > \SI{10}{\GeV{}}}$ muon in the next one, i.e., after the one where the corresponding \pp{} collision occurred.
Thus, the inclusion of this trigger increases the signal efficiency in the $0.4<\beta<0.8$ range.
The signal efficiency of the late-muon trigger is \SIrange[range-phrase = --]{2}{9}{\percent}.
The combined efficiency of all three triggers sums up to \SI{43}{\percent} on average.
It varies from \SI{28}{\percent} to \SI{62}{\percent} between the samples with $m=\SI{2000}{\GeV{}}$, $z=7$
and $m=\SI{500}{\GeV{}}$, $z=2$ MCPs, respectively.
 
To improve the description of the trigger simulation and to correct the trigger-efficiency values in the simulation,
parameterized corrections based on the time interval needed for MCPs to reach the RPC planes were applied to the
probability for MCPs to fire the RPC trigger.
 
The data selected by applying the logical OR of these three triggers are used in the analysis.
Only events recorded when all the ATLAS subdetectors were running at nominal conditions are used~\cite{DAPR-2018-01}.
 
\subsection{Candidate-track preselection}
\label{sec:CandTrackPreselAndIonisEstim}
 
A selected event is required to have at least one preselected candidate track.
Such a candidate track is required to be a `combined' muon, i.e., to be reconstructed by combining track segments in
the ID with those in the MS\@.
It also has to satisfy the `medium' criteria~\cite{MUON-2018-03},
to have ${\pT^{\mu}/z > \SI{50}{\GeV{}}}$, ${\pT/z > \SI{10}{\GeV{}}}$, and to fall within
the acceptance region of the TRT (${\abseta{} < 2.0}$).
For purposes described in \Sect{\ref{sec:TightAndFinalSels}}, this candidate also has to have
a defined \dEdx{} measurement in the pixel, TRT, and MDT.
 
To reduce the background of high-ionization signals from two or more particles firing the same TRT straws
or MDT tubes, such a candidate is required not to have any tracks with ${\pT{}/z > \SI{0.5}{\GeV{}}}$
within ${\Delta R = 0.01}$ reconstructed by the silicon detectors.
 
\subsection{Ionization estimators and tight/final selections}
\label{sec:TightAndFinalSels}

The average specific energy loss is described by the Bethe--Bloch formula~\cite{1930AnP...397..325B}.
Since a particle's energy loss increases quadratically with its charge,
an MCP would leave a very characteristic signature of high ionization in the detector.
Ionization losses in the sensitive elements are evaluated for the pixel, TRT, and MDT subdetector systems.
The pixel \dEdx{} is calculated from the truncated mean of the \dEdx{} values of the clusters associated with the track
by excluding the largest \dEdx{} measurement(s): one in case of three or four initial clusters,
or two in case of a larger number of initial clusters~\cite{SUSY-2016-03}.
IBL clusters with the overflow bit raised are also excluded from this calculation.
The truncated-mean approach allows large fluctuations in the \dEdx{} measurements to be reduced and improves the \dEdx{} resolution.
The TRT \dEdx{} is the truncated mean of the straw-level \dEdx{} estimates,
derived from the time interval when the signal remains above the low threshold.
Each drift tube of the MDT system provides a signal proportional to the charge from ionization
collected during \SI{18.5}{\ns} after the first electrons arrive at the wire;
a truncated mean of these measurements is treated as the MDT \dEdx{} estimator.
For the TRT and MDT, only one measurement per track is excluded as the final track-level distributions of
TRT or MDT \dEdx{} do not change significantly when more than one measurement is excluded.
Calibrations and corrections of the \dEdx{} estimators include removal of their dependencies:
\begin{itemize}
\item related to various detector effects: dependence on the number of hits,
radiation damage leading to run-by-run response differences for the pixel detector~\cite{ATLAS:2022pib},
occupancy for the pixel detector and TRT,
differences between the responses in the different detector sections for the MDT, etc.;
\item on geometrical quantities: pseudorapidity, distance between a particle track and an anode wire for the TRT and MDT.
The calibration is performed by assessing the difference between the \dEdx{} estimators in data and simulation
in a $\Zmm{}$ control sample and by reweighting the \dEdx{} estimators in bins of these geometrical quantities
in simulation based on the data distributions.
\end{itemize}
 
The significance of the \dEdx{} variable in each subdetector is defined by comparing the observed signal,
$\text{\dEdx{}}$, with the average value for a highly relativistic muon:
 
\begin{equation*}
S(\dEdx{}) = \frac{\text{\dEdx{}} - \langle\text{\dEdx{}}\rangle_{\mu}}{\sigma(\text{\dEdx{}})_{\mu}}.
\end{equation*}
 
Here $\langle\text{\dEdx{}}\rangle_{\mu}$ and $\sigma(\text{\dEdx{}})_{\mu}$
represent the mean and the root-mean-square width of the \dEdx{} distribution, respectively, for such muons in data
obtained by fitting the cores of the corresponding distributions with Gaussian functions.
To calculate these two parameters, a control sample of muons was obtained from $\Zmm{}$ events.
The muon selection is the same as in the analysis selection discussed in \Sect{\ref{sec:CandTrackPreselAndIonisEstim}}.
Additionally, muons are required to belong to an oppositely charged pair with its dimuon mass
between \SI{81}{\GeV{}} and \SI{101}{\GeV{}}, corresponding to detector resolution and the natural width of the $Z$~boson mass peak.
 
\begin{figure}[htb]
\begin{center}
\includegraphics[width=0.6\textwidth]{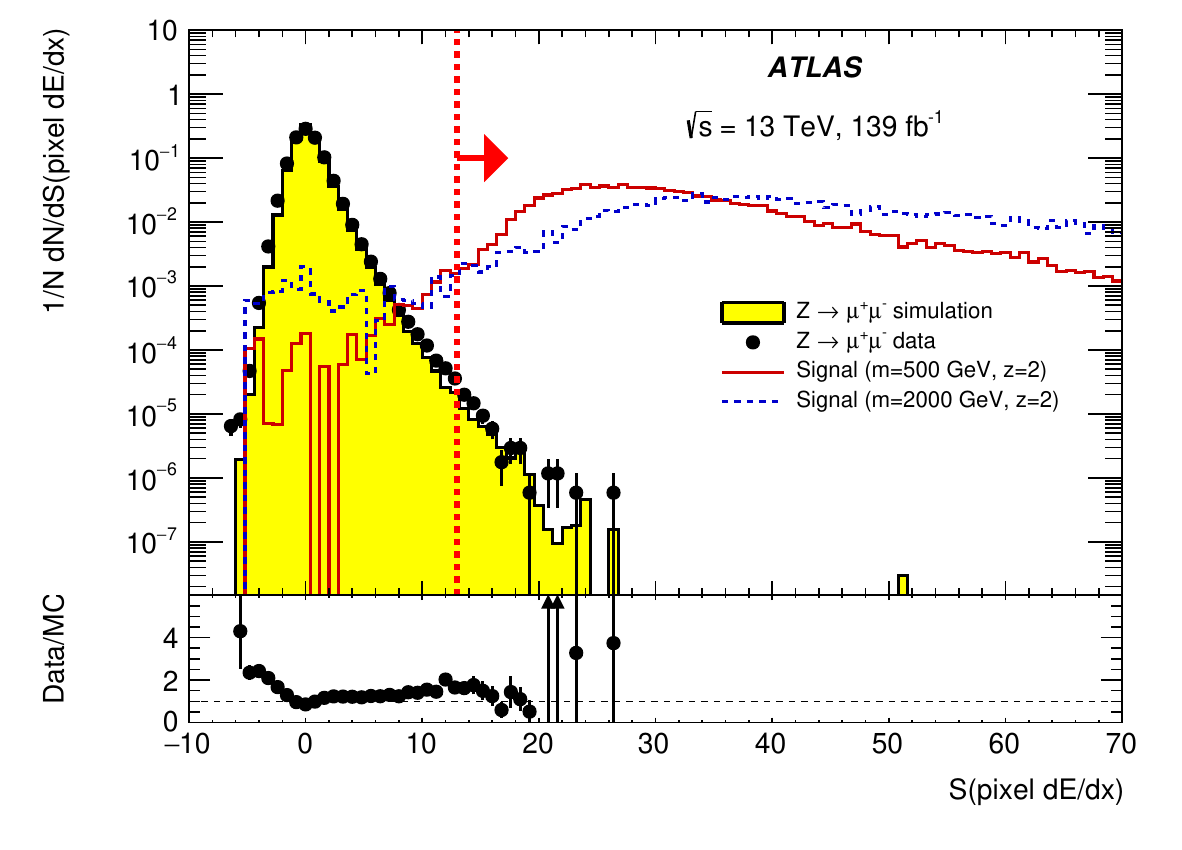}
\end{center}
\caption{Normalized distributions of the \dEdx{} significance in the pixel system,
$S$(pixel \dEdx{}), for muons from $\Zmm{}$ events (data and simulation) and for simulated MCPs
passing the preselection requirements.
Signal distributions are shown for ${z=2}$ and masses of $500$ and \SI{2000}{\GeV{}}.
The vertical dotted line indicates the threshold of the tight selection criterion.
A bin-by-bin ratio of $\Zmm{}$ distributions in data and simulation is shown on the lower panel.
The arrows in the ratio plot are for points that are outside the range.}
\label{fig:SPixeldEdx}
\end{figure}

As seen in \Fig{\ref{fig:SPixeldEdx}}, $S$(pixel \dEdx{}) is a powerful discriminator for ${z=2}$ particles.
The signal region of the tight selection for the ${z=2}$ category is defined by requiring $S$(pixel \dEdx{}) greater than $13$.
This tight selection requirement reduces the background contribution (mainly from the high-\pT{} muons)
by four orders of magnitude,
while keeping the signal efficiency above \SI{98}{\percent} relative to the previous selection step.
For ionization losses $8$ times greater than those for MIPs, the pixel readout saturates and the corresponding hits are not recorded,
which is why the same criterion cannot be used for the search for ${z>2}$ MCPs.
 
\begin{figure}[htb]
\begin{center}
\begin{subfigure}[b]{0.49\textwidth}
\centering
\includegraphics[width=\textwidth]{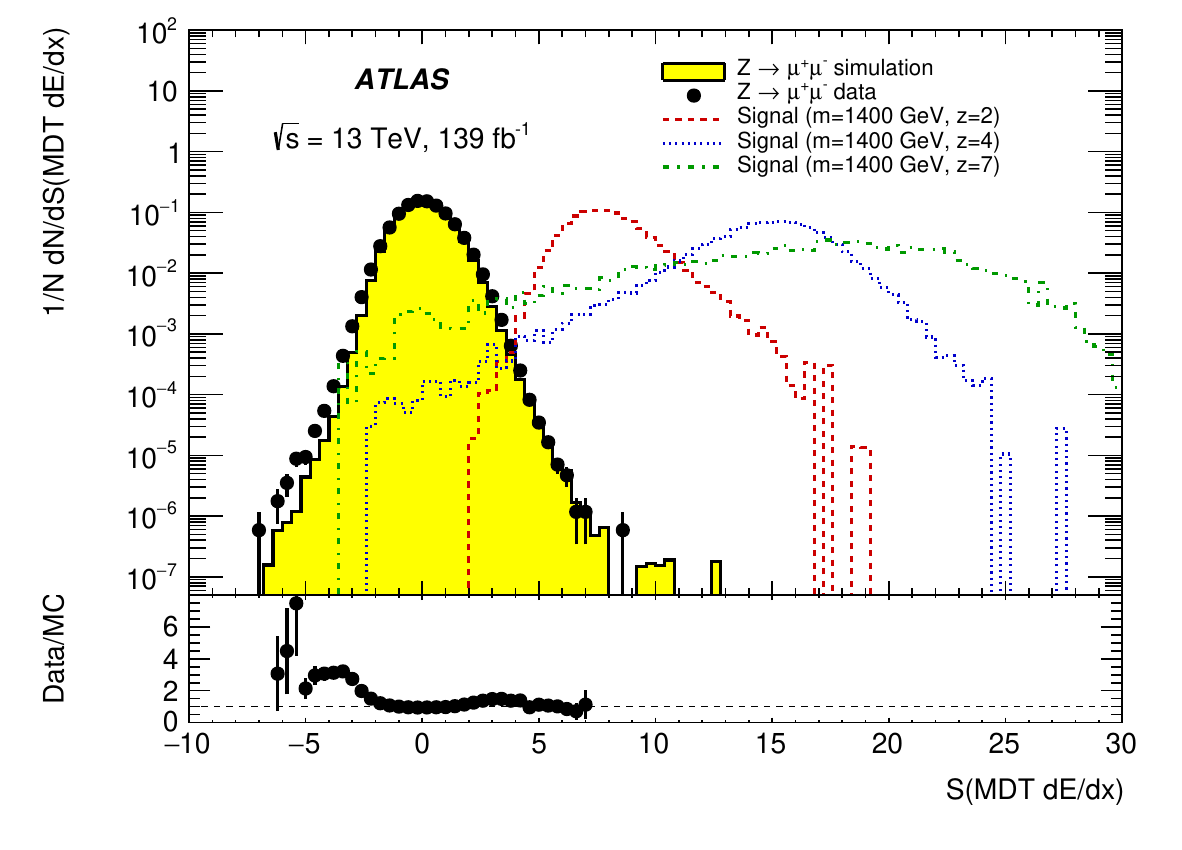}
\caption{}
\label{fig:dEdxSignificanceSeparation_SMdtDedx}
\end{subfigure}
\hfill
\begin{subfigure}[b]{0.49\textwidth}
\centering
\includegraphics[width=\textwidth]{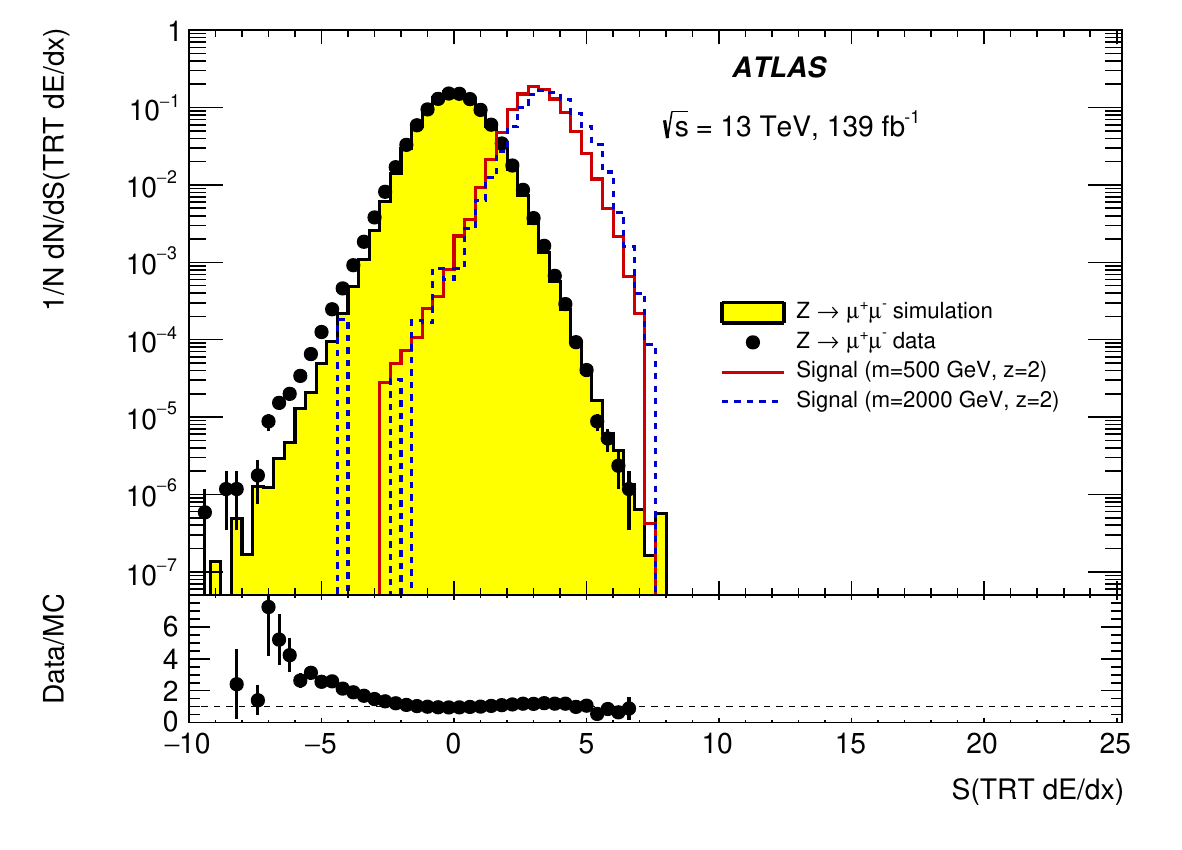}
\caption{}
\label{fig:dEdxSignificanceSeparation_STrtDedx}
\end{subfigure}
\hfill
\begin{subfigure}[b]{0.49\textwidth}
\centering
\includegraphics[width=\textwidth]{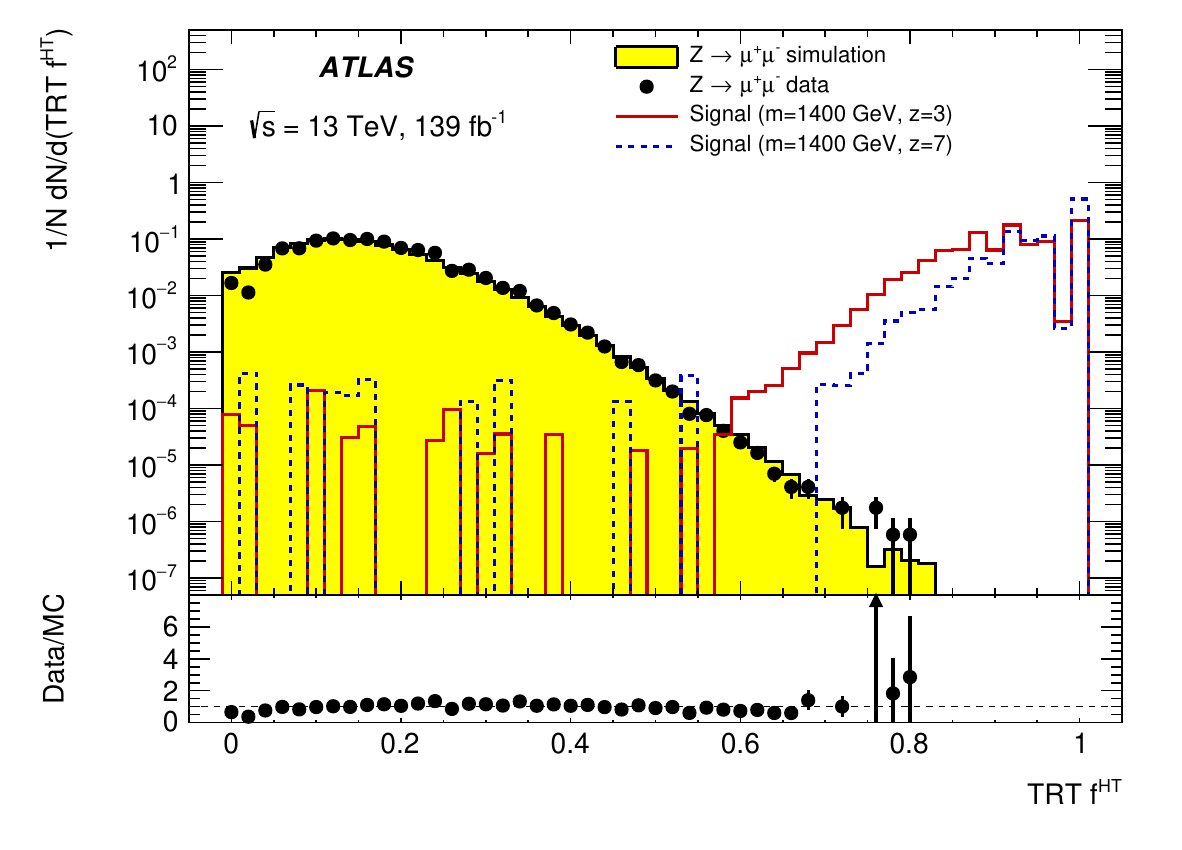}
\caption{}
\label{fig:dEdxSignificanceSeparation_TrtFht}
\end{subfigure}
\caption{Normalized distributions of the \dEdx{} significance
in \subref{fig:dEdxSignificanceSeparation_SMdtDedx} the MDT, $S$(MDT \dEdx{}),
\subref{fig:dEdxSignificanceSeparation_STrtDedx} the TRT, $S$(TRT \dEdx{}), and \subref{fig:dEdxSignificanceSeparation_TrtFht}
of TRT \fHT{} for muons from $\Zmm{}$ events (data and simulation) and for
simulated MCPs passing preselection requirements.
Bin-by-bin ratios of $\Zmm{}$ distributions in data and simulation are shown on the lower panels.
The arrow in the ratio plot in~\subref{fig:dEdxSignificanceSeparation_TrtFht} is for a point that is outside the range.}
\label{fig:dEdxSignificanceSeparation_SMdtAndSTrtAndTrtFht}
\end{center}
\end{figure}

In the final selection, $S$(MDT \dEdx{}), $S$(TRT \dEdx{}), and the fraction of HT TRT hits
(TRT \fHT{}, the number of HT hits divided by the number of LT hits on the track)
are used as additional discriminating variables to separate signal from background.
\Fig{\ref{fig:dEdxSignificanceSeparation_SMdtAndSTrtAndTrtFht}} shows the distributions of
these variables for muons from $\Zmm{}$ events compared with those expected from signal particles
with different charges (${z=2}$, $3$, $4$, and $7$) and masses ($500$, $1400$, and \SI{2000}{\GeV{}}).
It demonstrates that there is good separation between signal and background, which increases with increasing charge.
The $S$(MDT \dEdx{}) distribution broadens noticeably with charge because, relative to typical muons,
MCPs produce large number of $\delta$-rays, which give early-time hits.
As charge is measured in the first \SI{18.5}{\ns} of the signal, the ionization loss of $\delta$-rays is often
measured instead of the MCP's loss.
This gives large fluctuations in the total ionization measured along the track.
 
The detailed detector response to these high-charge particles may not be well
simulated due to a possible imperfect modelling of the saturation effects.
However, since the TRT and MDT do not lose signal at saturation,
their most probable \dEdx{} values are higher than those of ${z=2}$ particles.
Because of the conservative selections used at the final selection stage, the analysis is not sensitive
to the exact shape (or position) of the \dEdx{}-significance distributions for high-$z$ signals.
 
\begin{figure}[htb]
\begin{center}
\begin{subfigure}[b]{0.49\textwidth}
\centering
\includegraphics[width=\textwidth]{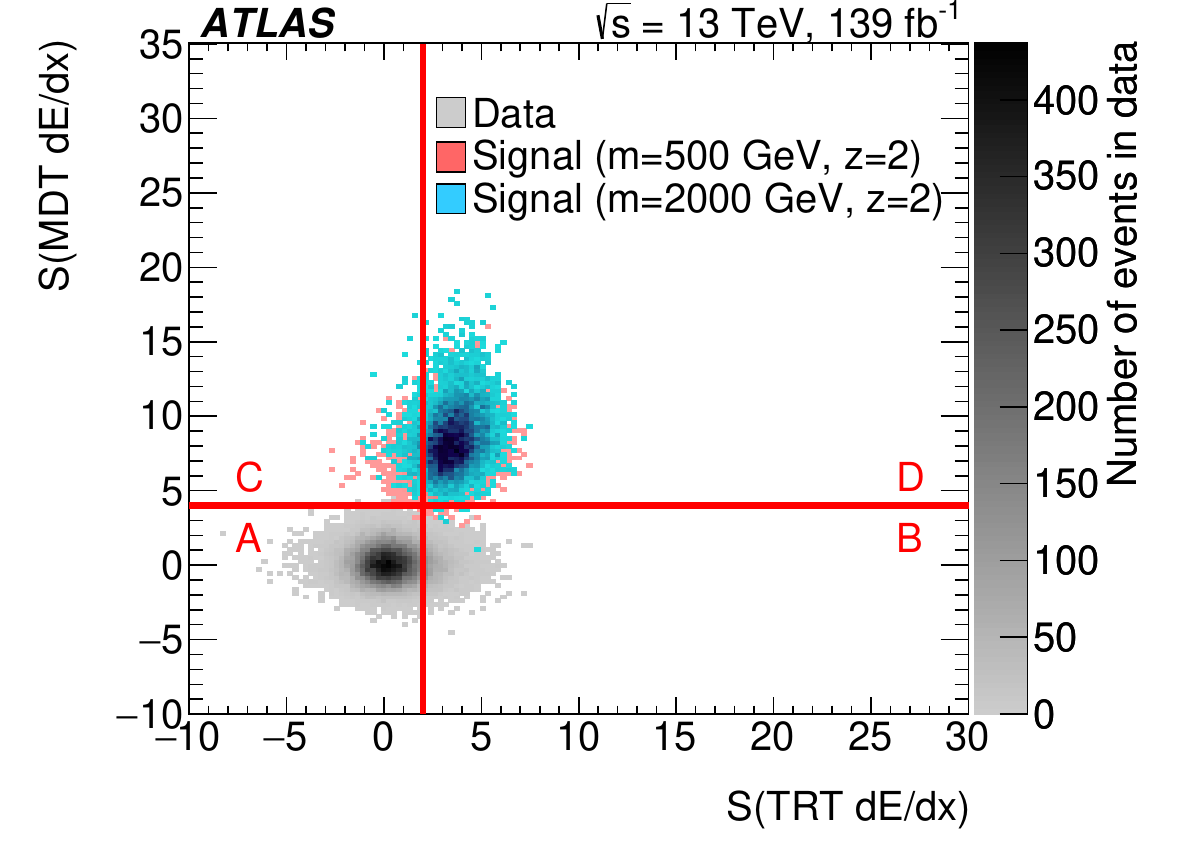}
\caption{}
\label{fig:AbcdPlanes_zEq2}
\end{subfigure}
\hfill
\begin{subfigure}[b]{0.49\textwidth}
\centering
\includegraphics[width=\textwidth]{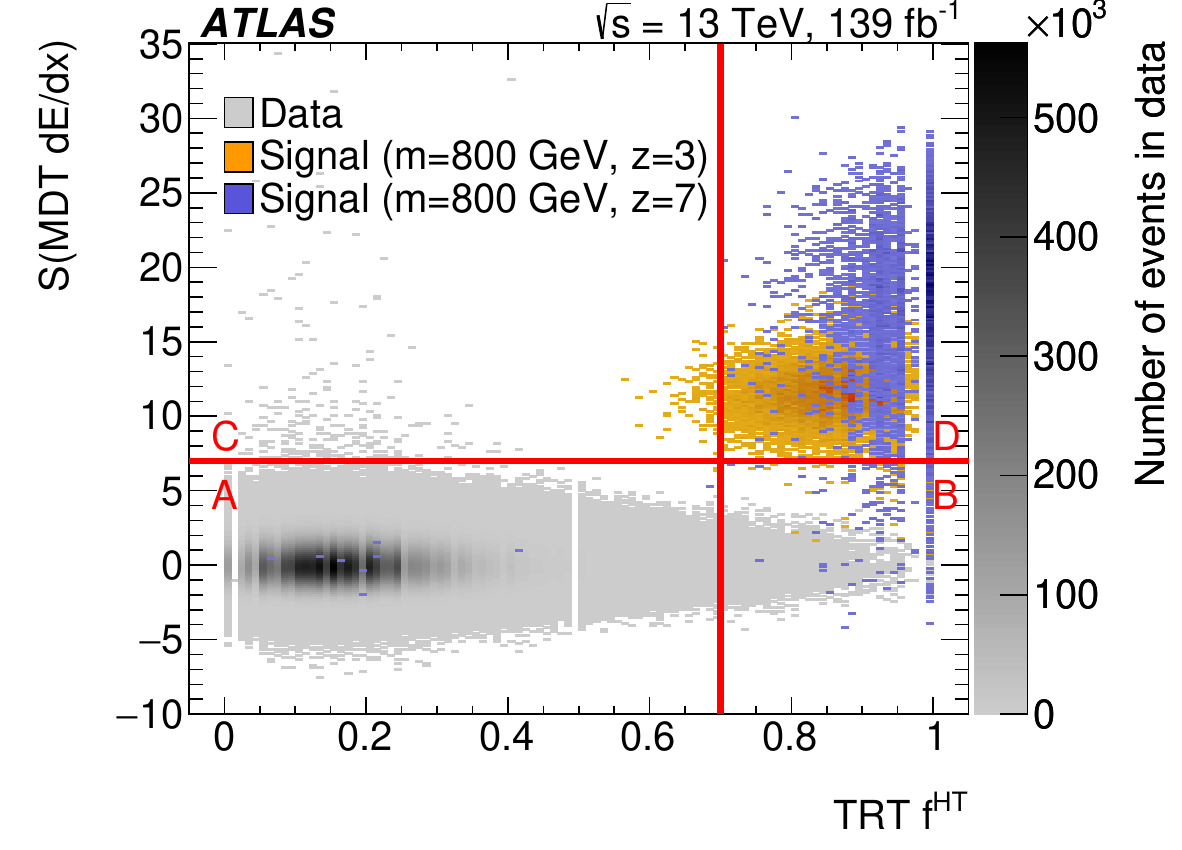}
\caption{}
\label{fig:AbcdPlanes_zGt2}
\end{subfigure}
\caption{\subref{fig:AbcdPlanes_zEq2} $S$(MDT \dEdx{}) versus $S$(TRT \dEdx{})
(used for the ${z=2}$ search) and \subref{fig:AbcdPlanes_zGt2} $S$(MDT \dEdx{}) versus TRT \fHT{}
(used for the ${z>2}$ search).
The distributions of the data and the simulated signal samples
(for charges ${z=2}$, $3$, and $7$, and masses of $500$, $800$, and \SI{2000}{\GeV{}}) are shown.
The signal distributions for the lowest MCP mass~\subref{fig:AbcdPlanes_zEq2} or charge~\subref{fig:AbcdPlanes_zGt2}
are partially covered by the other signal distributions.
D is the signal region and other regions are used to estimate the background contribution in that region (see text).
Regions $0.01 < $ TRT $\fHT{} < 0.02$, $0.49 < $ TRT $\fHT{} < 0.50$, and $0.98 < $ TRT $\fHT{} < 0.99$
in~\subref{fig:AbcdPlanes_zGt2} are underpopulated because TRT \fHT{} is a ratio of two integers belonging
to the range of \numrange[range-phrase = --]{0}{50}.}
\label{fig:AbcdPlanes}
\end{center}
\end{figure}

Two-dimensional distributions of $S$(MDT \dEdx{}) versus $S$(TRT \dEdx{})
and versus TRT \fHT{}
are shown for data and simulated signal events in \Fig{\ref{fig:AbcdPlanes}} for candidates
passing the tight selection for ${z=2}$ (\Fig{\ref{fig:AbcdPlanes_zEq2}}) and
the preselection for ${z>2}$ (\Fig{\ref{fig:AbcdPlanes_zGt2}}).
The two signal regions are defined by $S$(TRT \dEdx{}) $> 2$ and $S$(MDT \dEdx{}) $> 4$ for
candidates selected as ${z=2}$ and by TRT $\fHT{} > 0.7$ and $S$(MDT \dEdx{}) $> 7$
for candidates selected as ${z>2}$.
The choice of these criteria is discussed below.
The selection criteria were defined using simulated samples and $\Zmm{}$ data control samples
without examining the signal region in the data.
 
A summary of the offline-selection requirements (preselection, tight selection and final selection)
is presented in \Tab{\ref{tab:SummaryOfOfflineSelectionCriteria}}.
 
\begin{table}[htb]
\caption{Summary of the offline-selection requirements.}
\begin{center}
\scalebox{0.9}{
\begin{tabular}{l|c|c|c}
Search category 	& Preselection 							& Tight selection 		& Final selection \\
\hline
\multirow{4}{*}{$z=2$} 	& Combined muon with: 						& 				& Tightly selected candidate with: \\
& 								& Preselected candidate with 	& \\
& `medium' identification criteria, 				& $S$(pixel \dEdx{}) $> 13$ 	& $S$(TRT \dEdx{}) $> 2$, \\
& $\pT^{\mu}/z > \SI{50}{\GeV{}}$, 				& 				& $S$(MDT \dEdx{}) $> 4$ \\
\cline{1-1}\cline{3-4}
\multirow{4}{*}{$z>2$}	& $\pT/z > \SI{10}{\GeV{}}$, 					& \multirow{4}{*}{--}		& Preselected candidate with: \\
& $\abseta{} < 2.0$, 						& 				& \\
& no other particles with 					& 				& TRT $\fHT{} > 0.7$, \\
& $\pT{}/z > \SI{0.5}{\GeV{}}$ within $\Delta R = 0.01$ 	& 				& $S$(MDT \dEdx{}) $> 7$ \\
\end{tabular}
}
\end{center}
\label{tab:SummaryOfOfflineSelectionCriteria}
\end{table}
 
Among all simulated events triggered by the muon triggers, there are none where a reconstructed muon
matches the trigger object but MCP(s) in the signal region in the same event does (do) not.
 
\section{Signal efficiency}
\label{sec:sigEff}
 
\begin{figure}[htb]
\begin{center}
\begin{subfigure}[b]{0.49\textwidth}
\centering
\includegraphics[width=\textwidth]{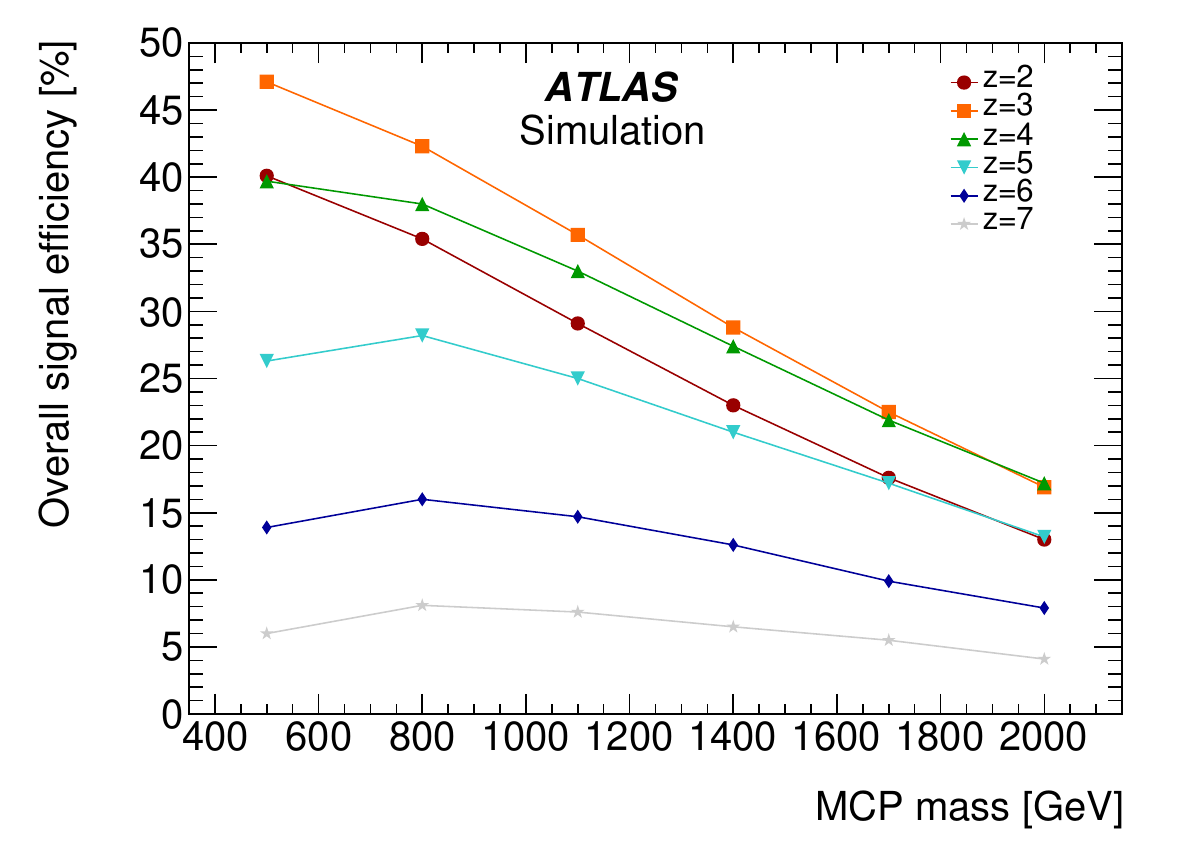}
\caption{}
\label{fig:EfficiencyTrends_vsMass}
\end{subfigure}
\hfill
\begin{subfigure}[b]{0.49\textwidth}
\centering
\includegraphics[width=\textwidth]{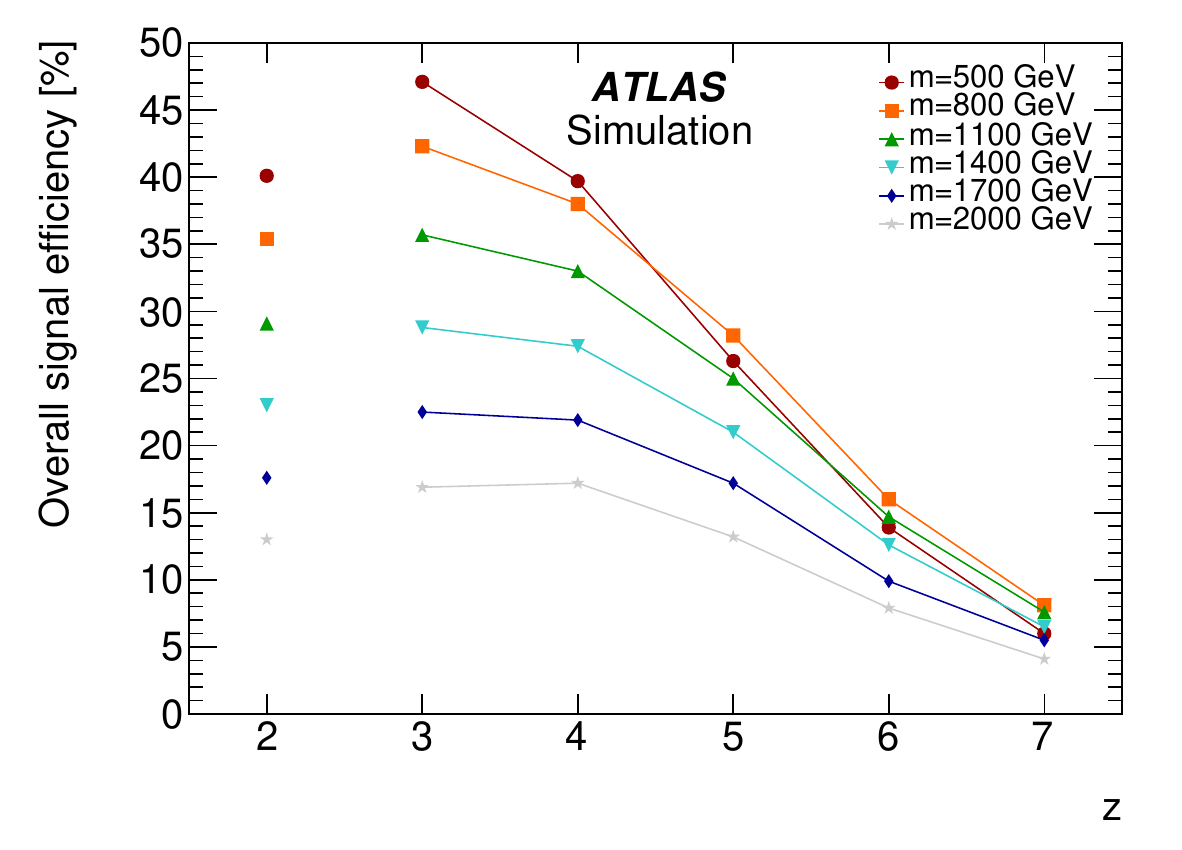}
\caption{}
\label{fig:EfficiencyTrends_vsCharge}
\end{subfigure}
\caption{The signal efficiencies for spin-\textonehalf{} MCPs with different charges and masses for the DY+PF production mode versus \subref{fig:EfficiencyTrends_vsMass} their mass and \subref{fig:EfficiencyTrends_vsCharge} their charge. Efficiency values for ${z=2}$ in~\subref{fig:EfficiencyTrends_vsCharge} are separated from the rest of the plot to reflect different signal selections used in the two search categories.  
}
\label{fig:EfficiencyTrends}
\end{center}
\end{figure}

The overall signal efficiency, which includes trigger, reconstruction and selection
efficiencies, is estimated from simulation.
Its values are shown in \Fig{\ref{fig:EfficiencyTrends}} for the signal samples used in this analysis.
The fractions of simulated signal events satisfying the cumulative selection requirements, as well as expected signal yields,
are given in \Tab{\ref{tab:EfficiencyCutflow}} for several benchmark points.
 
\begin{table}[htb]
\caption{Fractions of simulated signal events with at least one spin-\textonehalf{} MCP candidate, which satisfy the given requirements (including all previous selection requirements).
The uncertainties quoted are statistical only.
The expected signal yields, calculated as a product of theoretical cross-section, overall signal efficiency, and integrated luminosity, are shown in the rightmost column.}
 
\begin{center}
\begin{tabular}{|c|r|c c c|c|c|}
\hline
\multicolumn{2}{|c|}{Signal benchmark point} 	& Trigger 			& Candidate-track 		& Tight 			& Final 			& Expected signal	\\
\cline{1-2}
$z$	& Mass [\si{\GeV{}}]	 		& selection [\si{\percent}]	& preselection [\si{\percent}]	& selection [\si{\percent}]	& selection [\si{\percent}]	& yields [events]	\\
\hline
\hline
& \num{500} 	& \num{61.7+-0.2} & \num{42.7+-0.2}	& \num{42.5+-0.2} 	& \num{40.1+-0.2}	& \num{460} \\
$2$ 	& \num{1100} 	& \num{49.9+-0.2} & \num{30.8+-0.2}	& \num{30.6+-0.2} 	& \num{29.1+-0.2} 	& \num{5.3} \\
& \num{2000}	& \num{32.8+-0.2} & \num{13.7+-0.2}	& \num{13.6+-0.2} 	& \num{13.0+-0.2} 	& \num{0.026} \\
\hline
& \num{500} 	& \num{55.3+-0.2} & \num{39.9+-0.2}	& -- 			& \num{39.7+-0.2} 	& \num{2700} \\
$4$ 	& \num{1100}	& \num{50.5+-0.2} & \num{33.3+-0.2}	& -- 			& \num{33.1+-0.2} 	& \num{36} \\
& \num{2000}	& \num{35.0+-0.2} & \num{17.5+-0.2}	& -- 			& \num{17.2+-0.2} 	& \num{0.22} \\
\hline
& \num{500}	& \num{33.3+-0.2} & \;\:\num{6.2+-0.1}	& -- 			& \;\:\num{6.0+-0.1} 	& \num{2300} \\
$7$ 	& \num{1100} 	& \num{36.1+-0.2} & \;\:\num{8.1+-0.1}	& -- 			& \;\:\num{7.6+-0.1} 	& \num{49} \\
& \num{2000}	& \num{27.7+-0.2} & \;\:\num{4.7+-0.1}	& -- 			& \;\:\num{4.1+-0.1} 	& \num{0.31} \\
\hline
\end{tabular}
\end{center}
\label{tab:EfficiencyCutflow}
\end{table}
 
Several factors contribute to the efficiency dependencies on mass and charge.
For low masses, the ${\abseta{}<2.0}$ selection requirement and especially
the ${\pT{}/z}$ one are the main sources of efficiency loss.
For instance, the ${\pT{}/z}$ selection can be as high as approximately
$\pT{}>50\times 7 = \SI{350}{\GeV{}}$ (for ${z=7}$) and is the root cause of the efficiency drop at low masses.
For high masses, the reconstruction efficiency of muons is the primary reason for the reduction in efficiency.
Also, high ionization loss makes particles slow down: they may arrive at the MS too late to be recorded by the trigger
or lose all their kinetic energy before reaching the MS\@.
The charge dependence of the efficiency results from higher ionization
and the stricter ${\pT{}/z}$ requirement, which are augmented by the factors of $z^2$ and $z$, respectively.
Finally, large $\delta$-electron yields in the case of the highest-charged MCPs distorts the timing parameters of MDT hits from MCPs
leading to a smaller number of reconstructed combined muons. 
The main reason for the search to be limited to a maximum of ${z=7}$ is that MCPs with higher charges will fail to be
reconstructed as muons in at least \SI{95}{\percent} of cases.
 
\section{Expected background estimation}
\label{sec:bkgEstim}
 
The potential background mainly consists of muons with ionization randomly
fluctuating toward larger values due to detector occupancy effects (many particles losing
their energy in the same detector elements) and $\delta$-ray yields.
Radiation background and sporadic-noise events may also contribute to a large deviation of \dEdx{} measurements.
 
The expected background rate is estimated by using an ABCD method~\cite{EXOT-2012-10} in which the plane of two uncorrelated variables
is divided into regions A, B, C, and D as shown in \Fig{\ref{fig:AbcdPlanes}}.
Region D is defined as the signal region using the final selection cuts
on the $S$(TRT \dEdx{}) and $S$(MDT \dEdx{}) for the ${z=2}$ search and
on the TRT \fHT{} and $S$(MDT \dEdx{}) for the ${z>2}$ search, with regions A, B, and C as control regions.
 
The final selection cuts are set based on the following considerations:
\begin{itemize}
\item the number of expected background events in the D region should be kept minimal to provide more
stringent exclusion limits if nothing is observed in this region or to increase the excess significance in the alternative case;
\item the fraction of simulated signal events not satisfying at least one selection requirement should be kept
minimal ($\leq\SI{5}{\percent}$);
\item the numbers of data events in the control regions should remain statistically significant ($>5$ events) in order
to not introduce a large statistical uncertainty on the number of events expected in the D region from background processes.
\end{itemize}
 
If two or more candidates from the same event appear on the ABCD plane, only one is
retained and shown on the plane according to a `D-C-B-A-$\pT{}$' ranking to avoid double-counting events.
This ranking is first established by choosing an MCP candidate going from the most to the least populated region
in signal simulation: D$\rightarrow$C$\rightarrow$B$\rightarrow$A.
If there are two or more MCP candidates in the event and in the same region, the highest-$\pT{}$ one is chosen.
The fraction of events in data with two or more candidates on the ABCD plane in the same event
is \SI{0.01}{\percent} and \SI{0.06}{\percent} for the ${z=2}$ and ${z>2}$ cases, respectively.
 
The expected number of background events in the D region, $\ndata^{\textrm{D expected}}$, is estimated
from the numbers of observed data events in regions A, B, and C ($\ndata^{\textrm{A, B, C observed}}$):
 
\begin{equation}
\ndata^{\textrm{D expected}} = \frac{\ndata^{\textrm{B observed}}\times \ndata^{\textrm{C observed}}}{\ndata^{\textrm{A observed}}}.
\label{eq:regularAbcd}
\end{equation}
 
The expected background contributions to the D regions and quantities used for their calculation
are shown in \Tab{\ref{tab:yields_bkg}}.
Systematic uncertainties in these values are estimated according to the method discussed in \Sect{\ref{sec:uncertainties}}.
 
\begin{table}[htb]
\caption{Expected background contributions (in events) to the D regions in data for the ${z=2}$ and ${z>2}$ selections, as well as quantities used for their calculation.
The observed contributions are shown in the rightmost column.}
\begin{center}
\scalebox{0.9}{
\begin{tabular}{l| c c c c c}
Search category & $\ndata^{\textrm{A observed}}$ & $\ndata^{\textrm{B observed}}$ & $\ndata^{\textrm{C observed}}$ & $\ndata^{\textrm{D expected}}$ & $\ndata^{\textrm{D observed}}$ \\
& & & & & \\[-1em] 
\hline
${z=2}$ & \num{41674} 	  & \num{5024} 	& \num{13}  & $1.6 \pm 0.4 \stat{} \pm 0.5 \syst{}$       & \num{4} \\
${z>2}$ & \num{192036934} & \num{15004} & \num{441} & $0.034 \pm 0.002 \stat{} \pm 0.004 \syst{}$ & \num{0} \\
\end{tabular}
}
\end{center}
\label{tab:yields_bkg}
\end{table}
 
Since the average signal leak to regions A, B, and C is less than \SI{1}{\percent} of the entire signal yield,
the background contribution was estimated neglecting this contamination.
 
\section{Uncertainties in the background estimation and signal yield}
\label{sec:uncertainties}
 
Uncertainties in the background estimate, the signal selection efficiency, and the integrated luminosity affect
the sensitivity of the search for MCPs.
The contributions of these systematic uncertainties are described below.

\subsection{Background estimation uncertainty}
 
\begin{figure}[htb]
\begin{center}
\includegraphics[width=0.6\textwidth]{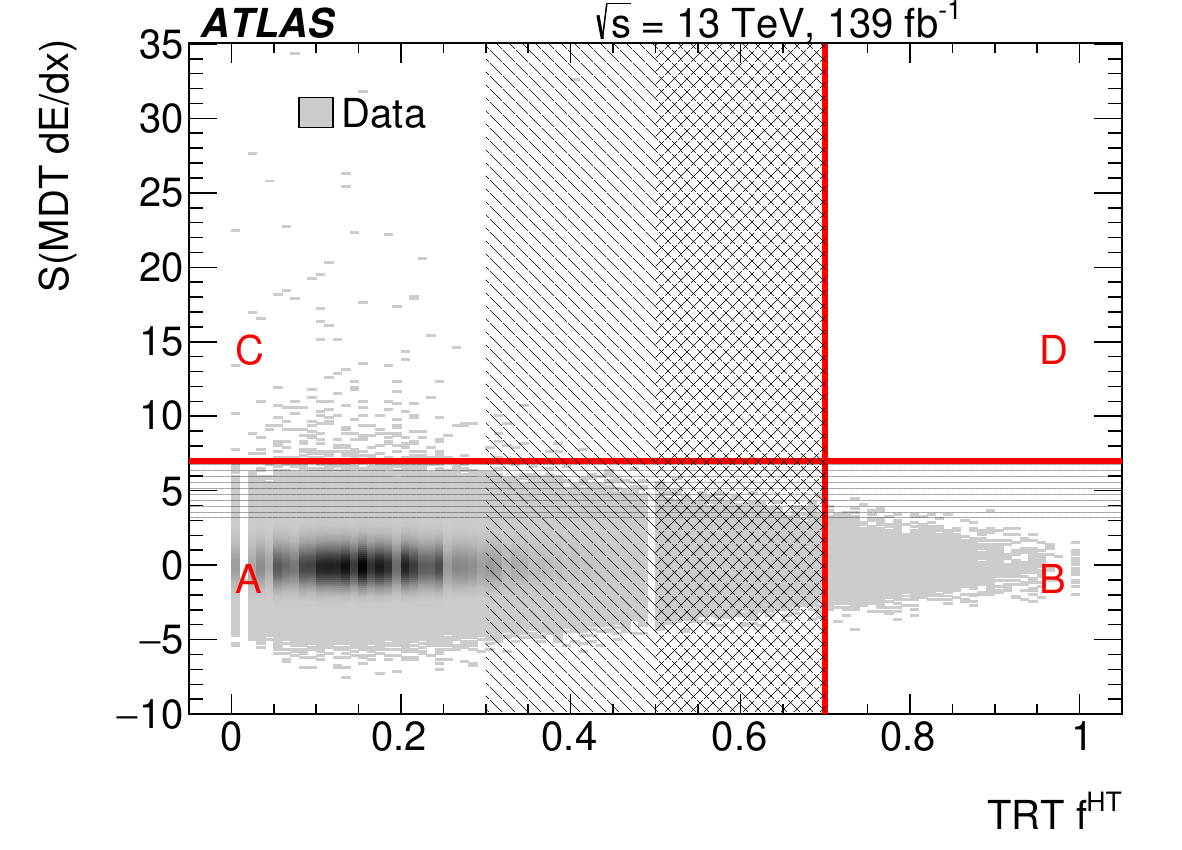}
\end{center}
\caption{ABCD plane for the ${z>2}$ case used to assess the systematic uncertainty in the expected number of background events.
Entries inside the `masked regions' (here with ${0.3 < \fHT{} < 0.7}$, ${0.5 < \fHT{} < 0.7}$, and ${3.0 < S(\text{MDT}\,\dEdx{}) < 7.0}$, shown by the different hatch styles) do not contribute to the background estimate.}
\label{fig:BkgUncertainty}
\end{figure}

To assess a systematic uncertainty in the expected number of background events, `masked regions'
are introduced in the ABCD planes, and then the background estimate is recalculated
for several masked-region choices using the method described in \Sect{\ref{sec:bkgEstim}}
while removing the entries in the masked regions from the calculation of the background estimate.
When the two variables on an ABCD plane depend on each other, the fraction of entries removed
in the A region is different from the same number for the B (or C) region.
This would lead to a different number of events expected in the D region compared to the original estimation.
The masked regions used are:
$S_{\textrm{lower}}^{\textrm{TRT}} < S$(TRT \dEdx{}) $< 2.0$ with
$S_{\textrm{lower}}^{\textrm{TRT}} = 0.8, 1.0, \ldots, 1.8$ and
$S_{\textrm{lower}}^{\textrm{MDT}} < S$(MDT \dEdx{}) $< 4.0$ with
$S_{\textrm{lower}}^{\textrm{MDT}} = 0.5, 1.0, \ldots, 3.5$ for the ${z=2}$ case, and
$\ensuremath{f^{\textrm{HT}}_{\textrm{lower}}} < \fHT{} < 0.7$ with
$\ensuremath{f^{\textrm{HT}}_{\textrm{lower}}} = 0.2, 0.3, \ldots, 0.6 $ and
$S_{\textrm{lower}}^{\textrm{MDT}} < S$(MDT \dEdx{}) $< 7.0$ with
$S_{\textrm{lower}}^{\textrm{MDT}} = 1.0, 2.0, \ldots, 6.0$ for the ${z>2}$ case.
Examples of masked regions with $\ensuremath{f^{\textrm{HT}}_{\textrm{lower}}} = 0.3$, $0.5$,
and $S_{\textrm{lower}}^{\textrm{MDT}}=3.0$ on the ${z>2}$ ABCD plane are shown in \Fig{\ref{fig:BkgUncertainty}}.
Assuming there are $\ndata^{\textrm{A observed, masked}}$, $\ndata^{\textrm{B observed, masked}}$, and
$\ndata^{\textrm{C observed, masked}}$ events in the masked areas of regions A, B, and C, respectively,
the recalculated background estimates are given by
 
\begin{equation*}
\frac{\left( \ndata^{\textrm{B observed}} - \ndata^{\textrm{B observed, masked}}\right)\times \ndata^{\textrm{C observed}}}{\ndata^{\textrm{A observed}} - \ndata^{\textrm{A observed, masked}}}
\end{equation*}
 
for horizontal masked regions and by
 
\begin{equation*}
\frac{\ndata^{\textrm{B observed}}\times\left(\ndata^{\textrm{C observed}} - \ndata^{\textrm{C observed, masked}}\right)}{\ndata^{\textrm{A observed}} - \ndata^{\textrm{A observed, masked}}}
\end{equation*}
 
for vertical masked regions.

The maximum differences (calculated over all the masked regions examined) between a new background expectation in the D
region and the nominal one are \SI{31}{\percent} ($0.5$~events) and \SI{12}{\percent} ($4 \times 10^{-3}$~events)
for the ${z=2}$ and ${z>2}$ cases, respectively, and are treated as systematic uncertainties in
the estimation of the expected background yield.
 
\begin{figure}[htb]
\begin{center}
\begin{subfigure}[b]{0.49\textwidth}
\centering
\includegraphics[width=\textwidth]{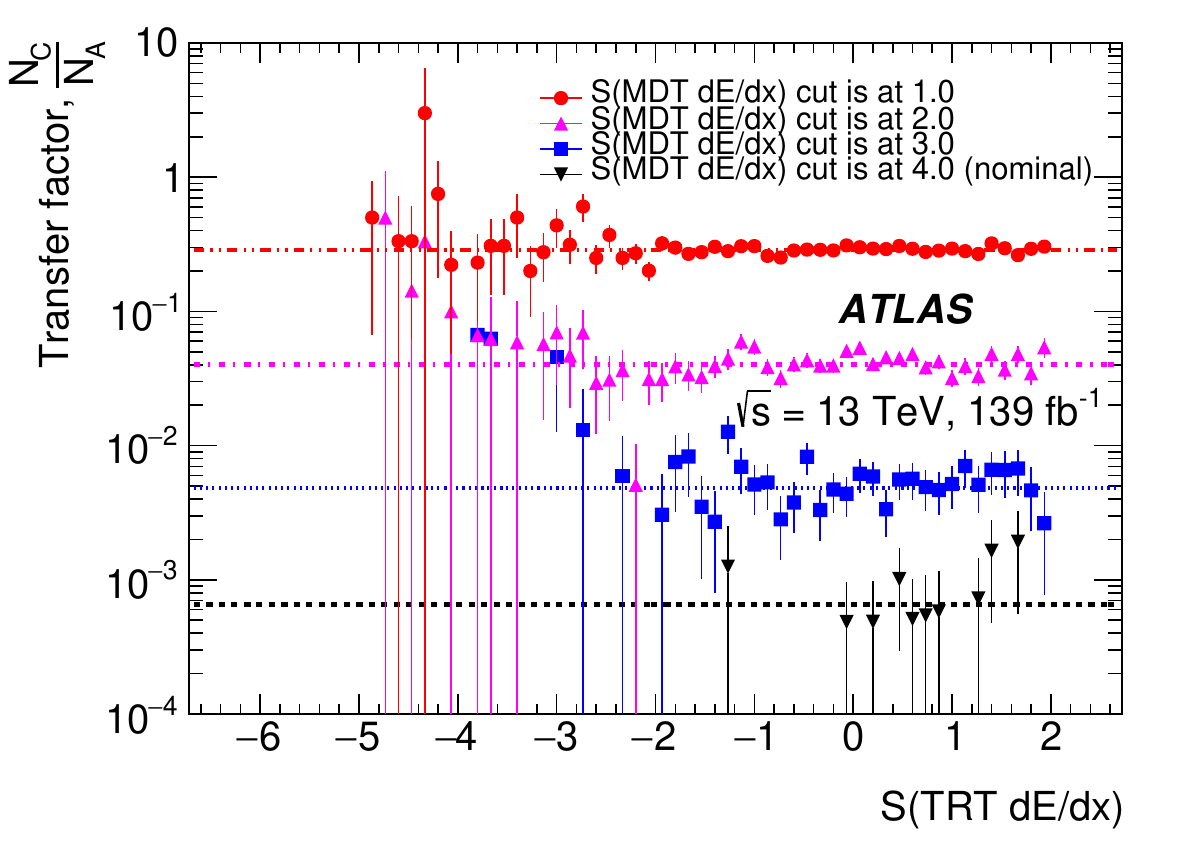}
\caption{}
\label{fig:transferFactorVsXaxisQuantity_zEq2}
\end{subfigure}
\hfill
\begin{subfigure}[b]{0.49\textwidth}
\centering
\includegraphics[width=\textwidth]{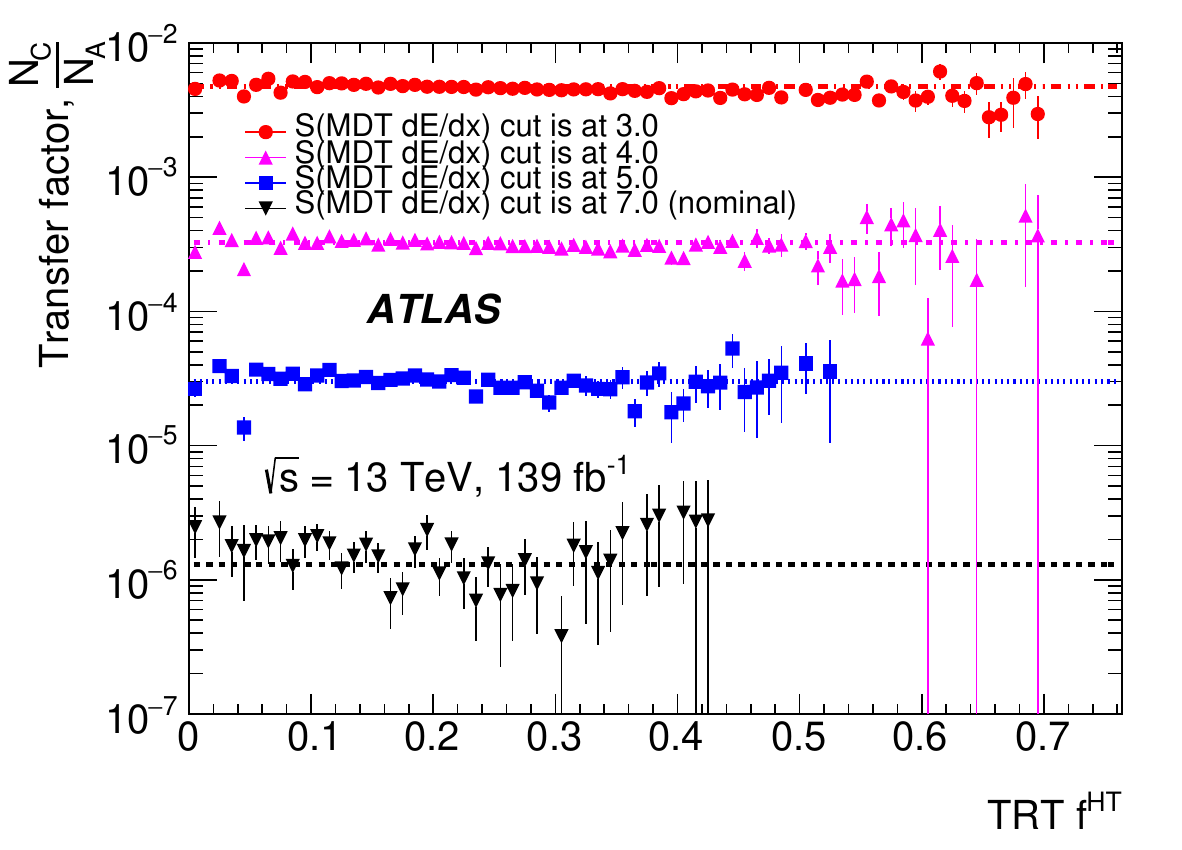}
\caption{}
\label{fig:transferFactorVsXaxisQuantity_zGt2}
\end{subfigure}
\caption{Transfer factor versus \subref{fig:transferFactorVsXaxisQuantity_zEq2} $S$(TRT \dEdx{}) (for ${z=2}$)
and \subref{fig:transferFactorVsXaxisQuantity_zGt2} TRT \fHT{} (for ${z>2}$) for several cuts on the
MDT-\dEdx{} significance. The horizontal lines of different styles represent constant fits to the respective distributions.}
\label{fig:transferFactorVsXaxisQuantity}
\end{center}
\end{figure}

The absence of correlations between ABCD-plane quantities (and quantities used at the earlier selection stages)
was discussed in Ref.~\cite{EXOT-2017-13}.
An additional check is performed to ensure \Eqn{(\ref{eq:regularAbcd})} is applicable for the data distributions on both ABCD planes.
The `transfer factor', $\frac{\ndata^{\text{C observed}}}{\ndata^{\text{A observed}}}$,
is plotted against $S$(TRT \dEdx{}) (for ${z=2}$) and TRT \fHT{} (for ${z>2}$).
To eliminate low-statistics effects, the cuts on the MDT-\dEdx{} significance
on the ABCD planes are iteratively relaxed from $4$ to $3$, $2$, and $1$ for the ${z=2}$ case
and from $7$ to $5$, $4$, and, finally, to $3$ for the ${z>2}$ case.
The resulting plots are shown in \Fig{\ref{fig:transferFactorVsXaxisQuantity}}.
There is no statistically significant evidence that the transfer factor increases
with $S$(TRT \dEdx{}) or \fHT{} in any of these scenarios, which suggests there is no additional source of systematic uncertainty.
Finally, possible contributions from second-order effects into the overall background distribution are expected
to be negligible and hence should not introduce correlations in the ABCD planes.
Considering how rare such events are, it is not practical to simulate a statistically significant sample.
 
\subsection{Signal efficiency uncertainty}
 
Several sources of systematic uncertainty in the signal efficiency are considered.
The most significant uncertainties are those due to some mismodelling of online
(trigger) and offline ($\dEdx{}$ estimators etc.) quantities.
 
The uncertainty due to the mismodelling of the offline quantities is evaluated by varying
the selection requirements.
Several considerations motivate these variations.
The uncertainty in the amount of material in front of the MS, which is about \SI{1}{\percent}~\cite{PERF-2014-05},
propagates into an uncertainty in the selection efficiency due to the slowing down of particles.
It is estimated by varying the $\pT^{\mu}/z$ and $\pT{} / z$ requirements by $\pm\SI{3}{\percent}$.
 
The areas around the peaks of the \dEdx{} distributions for the $\Zmm{}$ process are the most important when determining
the signal efficiency uncertainty as the quality of \dEdx{} modelling in this region is of most relevance for MCPs.
Slight disagreements between the shapes of the distributions in $\Zmm{}$ events
in data and simulation are accounted for by the variations of the signal selection criteria.
The values of these variations are obtained by averaging the bin-by-bin ratios of $\Zmm{}$ yields in data to those in
simulation (see \Figs{\ref{fig:SPixeldEdx}}{\ref{fig:dEdxSignificanceSeparation_SMdtAndSTrtAndTrtFht}}) in the cores
of the corresponding distributions (within ${\pm3}$ standard deviations of the mean of each distribution).
The selection criteria were varied as follows:
$S$(pixel \dEdx{}) by $\pm\SI{14}{\percent}$ ($13\rightarrow13\times(1\pm0.14)$),
\fHT{} by $\pm\SI{12}{\percent}$ ($0.7\rightarrow0.7\times(1\pm0.12)$),
$S$(TRT \dEdx{}) by $\pm\SI{5}{\percent}$, and $S$(MDT \dEdx{}) by $\pm\SI{6}{\percent}$.
The total systematic uncertainties in the efficiency
arising from these variations range between \SI{0.3}{\percent} and \SI{5}{\percent}, where the largest
uncertainty corresponds to lower-mass ${z=3}$ signal samples, which are fairly sensitive to
the \fHT{} variation.
A dedicated analysis of shape of high-$z$-particle clusters and multiplicity of such clusters
in the pixel detector did not reveal any additional source of systematic uncertainty.
 
The uncertainty in the trigger efficiency also has several sources, including
an uncertainty in the muon-trigger efficiency (\SI{3}{\percent} on average), accounting for differences between
triggering on the muons in data and simulation,
and an uncertainty in the \MET{} trigger efficiency (\SI{5}{\percent} on average).
The overall trigger-efficiency uncertainty for a particular simulated signal sample is calculated
by accounting for the fractions of events in which each trigger fired.
 
The uncertainty in the muon-trigger efficiency is a $\beta$-dependent uncertainty originating from uncertainties
in the modelling of the muon-trigger timing for particles with ${\beta < 1}$ for the traditional single-muon trigger and
the late-muon trigger.
To assess the uncertainty, the parameters of the trigger-efficiency corrections were varied.
The $\beta$ value of particles was varied between the ones measured in the ID and in the MS from
the hypothesized mass and measured momentum.
The time interval needed for a signal particle to reach the RPC trigger planes was varied by
the root-mean-square width of the timing distribution for muons measured in the full $\Zmm{}$ data sample.
The uncertainty, assessed as the maximum relative difference between the nominal efficiency values and
those obtained after the variations, averages to \SI{1}{\percent}.
For the TGC trigger, no mismatch between the timing distributions in data and simulation was observed;
therefore no uncertainty was assigned to the trigger efficiency.
An increase in the trigger efficiency due to larger ionization losses of MCPs compared to muons is not taken into account,
resulting in a conservative estimate of trigger efficiency for muon triggers.
 
The uncertainty in the \MET{} trigger efficiency depends on the accuracy of modelling the \MET{} turn-on curve,
assessed using the offline \MET{} spectra (in events triggered exclusively by the \MET{} trigger),
varied to account for any possible uncertainties in the \MET{} term~\cite{SUSY-2016-32}.
In addition, since the efficiencies of both the \MET{} trigger and late-muon trigger depend on an extra jet,
which relies on the parton-shower modelling, the QCD scale used in the simulation was varied by doubling and halving
the renormalization and factorization scales, and an additional \SI{2}{\percent}
was added to the uncertainty in the efficiency of these two types of triggers.
 
Contributions from the separate sources of the most significant systematic uncertainties in
the signal selection efficiency, as well as the resulting values of overall systematic uncertainties
calculated as the sum in quadrature of all the individual contributions, are shown in \Tab{\ref{tab:EffUnc}} for several benchmark points.
 
\begin{table}[htb]
\caption{The most significant individual contributions (in \si{\percent}) to the overall systematic uncertainties in the signal selection efficiency, as well as the resulting values of the relative uncertainties (rightmost column).}
\begin{center}
\begin{tabular}{|c|r|c c c|c|}
\hline
\multicolumn{2}{|c|}{Signal benchmark point} & Data--simulation            & Trigger                      & \multirow{2}{*}{Tracking [\si{\percent}]} & Overall uncertainty in \\
\cline{1-2}
$z$     & Mass [\si{\GeV{}}]                 & comparison [\si{\percent}]  & efficiency [\si{\percent}]   &                                           & the selection efficiency [\si{\percent}]  \\
\hline
\hline
& 500                                & 1                           & 1                        & 0.6                                       & 2 \\
2 	& 1400                               & 1                           & 2                        & 2                                         & 3 \\
& 2000                               & 1                           & 2                        & 3                                         & 4 \\
\hline
& 500                                & 1                           & 1                        & 0.4                                       & 2 \\
4 	& 1400                               & 0.3                         & 2                        & 0.7                                       & 2 \\
& 2000                               & 0.3                         & 2                        & 1                                         & 3 \\
\hline
& 500                                & 3                           & 2                        & 0.3                                       & 4 \\
7	& 1400                               & 0.9                         & 2                        & 0.5                                       & 4 \\
& 2000                               & 1                           & 3                        & 0.7                                       & 5 \\
\hline
\end{tabular}
\end{center}
\label{tab:EffUnc}
\end{table}
 
Additionally, a \SI{1.7}{\percent} uncertainty was assigned to the integrated luminosity.
This uncertainty is derived, following a methodology similar to that detailed in Ref.~\cite{ATLAS-CONF-2019-021},
from a calibration of the luminosity scale using $x$--$y$ beam-separation scans.
 
\section{Results}
 
Only four events with MCP candidates were found in the data in the ${z=2}$ MCP search,
and none were found in the ${z>2}$ search.
All four events are very close to the boundaries between the D and C (and D and B) regions.
These results are consistent with the expectations of $1.6 \pm 0.4 \stat{} \pm 0.5 \syst{}$ and
$0.034 \pm 0.002 \stat{} \pm 0.004 \syst{}$ background events for the ${z=2}$ and ${z>2}$ search categories,
respectively (a $p_0$ value of $0.06$, corresponding to the $Z$-significance\footnote{The relation
between a $Z$-significance and a $p_0$ value is given by $Z = \Phi^{-1}(1-p_0)$, where $\Phi^{-1}$ is
the inverse of the cumulative distribution for a unit Gaussian function.} of $1.5$,
was obtained for the ${z=2}$ search category).
 
Cross-section limits are computed within the \textsc{RooStats} framework~\cite{2010acat.confE..57M} using
the CL$_\textrm{s}$ method~\cite{Read:2002hq} to discriminate between the background-only hypothesis and
the signal-plus-background hypothesis.
Exclusion limits are computed for various MCP scenarios.
The signal selection efficiency, luminosity, expected and observed numbers of events and their uncertainties,
handled as nuisance parameters, are taken as input for pseudo-experiments,
resulting in an observed limit at \SI{95}{\percent} confidence level (CL).
 
\begin{figure}[htb]
\begin{center}
\includegraphics[width=0.6\textwidth]{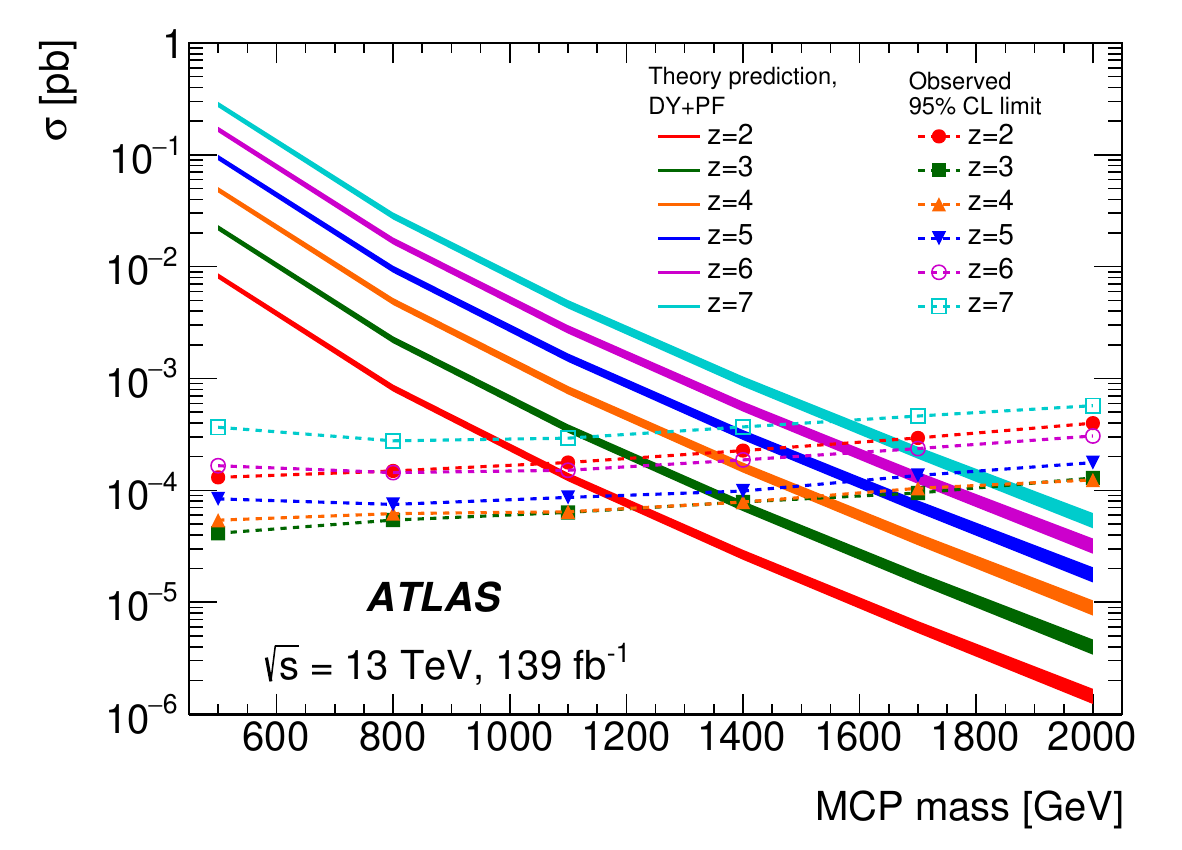}
\end{center}
\caption{Observed \SI{95}{\percent} CL cross-section upper limits and theoretical cross-sections as functions of the muon-like spin-\textonehalf{} MCP's mass for the DY+PF production mode. Theoretical cross-section values are computed at \LO{} and the uncertainty bands correspond to the PDF uncertainties.}
\label{fig:XSectionsOrTheirLimitsVsMass}
\end{figure}

The measurement excludes the DY+PF mode of muon-like MCP pair production over wide ranges of masses.
\Fig{\ref{fig:XSectionsOrTheirLimitsVsMass}} summarizes
the observed \SI{95}{\percent} CL cross-section limits as a function of mass for all the MCP charges studied and compares them
with those predicted by the DY+PF mode.
 
\begin{figure}[htb]
\begin{center}
\includegraphics[width=0.6\textwidth]{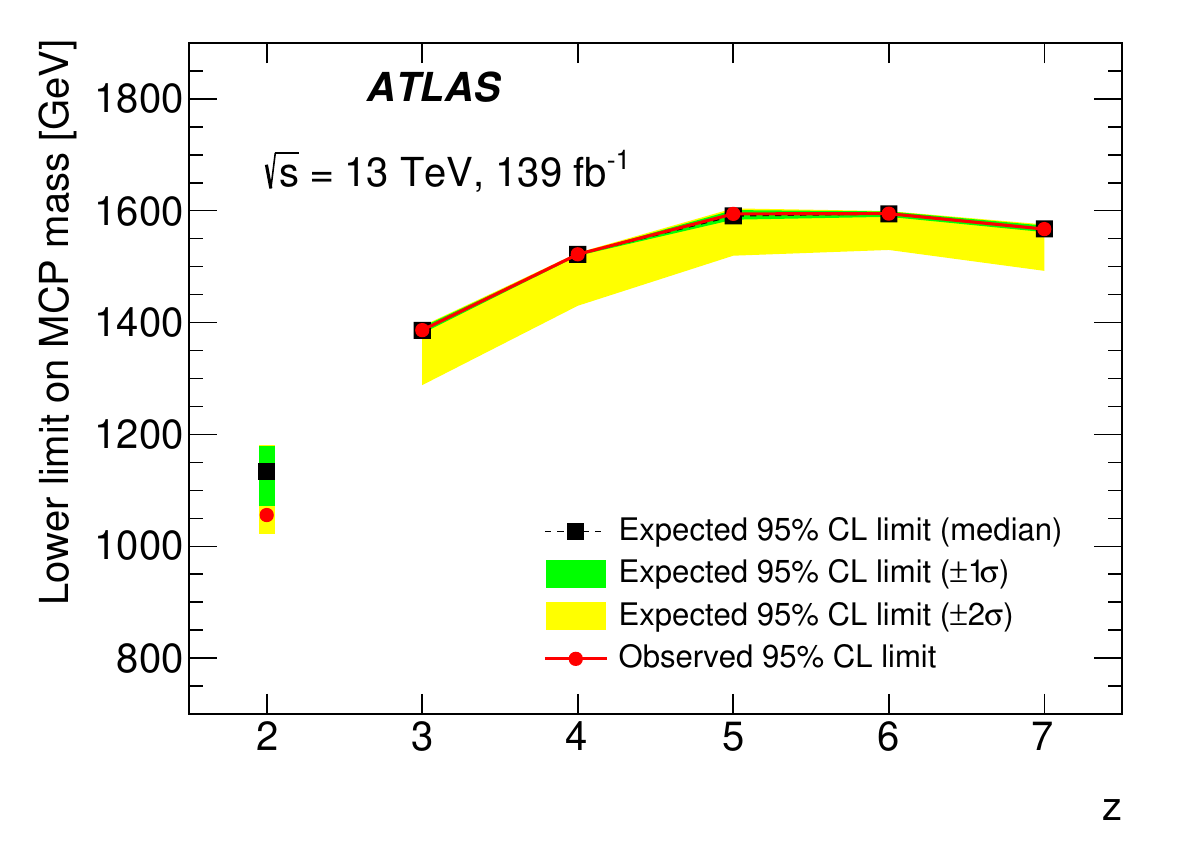}
\end{center}
\caption{Observed and expected \SI{95}{\percent} CL lower mass limits of MCPs versus charge.
The separation of limits between ${z=2}$ and ${z=3}$ is due to different selections used to identify MCPs with the respective charges.}
\label{fig:MassExclusion}
\end{figure}

The mass limits are obtained from the intersection of the observed cross-section limits
and the theoretical cross-section values not accounting for theoretical uncertainties.
For this production mode, the cross-section limits are transformed into
mass exclusion regions from \SI{500}{\GeV{}} up to the values in \Tab{\ref{tab:lowerMassLimits}}.
MCPs with mass values $ \SI{50}{\GeV{}} < m < \SI{500}{\GeV{}}$ were already excluded in the previous search~\cite{EXOT-2017-13}.
Observed and expected mass limits are compared in \Fig{\ref{fig:MassExclusion}}.
 
\begin{table}[htb]
\caption{Observed \SI{95}{\percent} CL lower mass limits of muon-like MCPs for the DY+PF production mode.}
\begin{center}
\begin{tabular}{c|c c c c c c}
$z$                            & 2    & 3    & 4    & 5    & 6    & 7    \\
\hline
Lower mass limit [\si{\GeV{}}] & 1060 & 1390 & 1520 & 1590 & 1600 & 1570 \\
\end{tabular}
\end{center}
\label{tab:lowerMassLimits}
\end{table}
 
Recently, an excess of events in a signal region in the ATLAS search for heavy long-lived $z=1$ particles
identifiable by their unusually large pixel \dEdx{} values~\cite{ATLAS:2022pib} was observed.
Two of these observed events feature candidates with pixel \dEdx{} values compatible with those satisfying
the ${z=2}$ tight-selection requirement in the current analysis, but not ending up in the corresponding signal region.
A dedicated check was performed to understand the reason for this.
It was demonstrated that neither of the two candidates have high enough ionization loss in TRT or MDT
to make it into the signal region -- in fact, both of them belong to the A control region (see \Fig{\ref{fig:AbcdPlanes_zEq2}}).

\FloatBarrier
 
\section{Conclusion}
 
This Letter reports on a search for long-lived multi-charged particles
produced in proton--proton collisions with the ATLAS detector at the LHC\@.
The search uses a data sample with a centre-of-mass energy of ${\rts=\SI{13}{\TeV{}}}$
and an integrated luminosity of \SI{139}{\ifb}.
Muon-like particles are searched for with electric
charges from ${|q|=2e}$ to ${|q|=7e}$ penetrating the full ATLAS detector
and producing anomalously high ionization signals in multiple detector elements.
No statistically significant evidence of such particles is observed.
Upper limits are derived on the cross-sections using a Drell--Yan + photon fusion production mode and
exclude muon-like multi-charged particles with masses between \SI{500}{\GeV{}}
and \SIrange[range-phrase = --]{1060}{1600}{\GeV{}}.
This production mode is used in an ATLAS search for multi-charged particles for the first time.
\FloatBarrier

\section*{Acknowledgements}


We thank CERN for the very successful operation of the LHC, as well as the
support staff from our institutions without whom ATLAS could not be
operated efficiently.
 
We acknowledge the support of
ANPCyT, Argentina;
YerPhI, Armenia;
ARC, Australia;
BMWFW and FWF, Austria;
ANAS, Azerbaijan;
CNPq and FAPESP, Brazil;
NSERC, NRC and CFI, Canada;
CERN;
ANID, Chile;
CAS, MOST and NSFC, China;
Minciencias, Colombia;
MEYS CR, Czech Republic;
DNRF and DNSRC, Denmark;
IN2P3-CNRS and CEA-DRF/IRFU, France;
SRNSFG, Georgia;
BMBF, HGF and MPG, Germany;
GSRI, Greece;
RGC and Hong Kong SAR, China;
ISF and Benoziyo Center, Israel;
INFN, Italy;
MEXT and JSPS, Japan;
CNRST, Morocco;
NWO, Netherlands;
RCN, Norway;
MEiN, Poland;
FCT, Portugal;
MNE/IFA, Romania;
MESTD, Serbia;
MSSR, Slovakia;
ARRS and MIZ\v{S}, Slovenia;
DSI/NRF, South Africa;
MICINN, Spain;
SRC and Wallenberg Foundation, Sweden;
SERI, SNSF and Cantons of Bern and Geneva, Switzerland;
MOST, Taiwan;
TENMAK, T\"urkiye;
STFC, United Kingdom;
DOE and NSF, United States of America.
In addition, individual groups and members have received support from
BCKDF, CANARIE, Compute Canada and CRC, Canada;
PRIMUS 21/SCI/017 and UNCE SCI/013, Czech Republic;
COST, ERC, ERDF, Horizon 2020, ICSC-NextGenerationEU and Marie Sk{\l}odowska-Curie Actions, European Union;
Investissements d'Avenir Labex, Investissements d'Avenir Idex and ANR, France;
DFG and AvH Foundation, Germany;
Herakleitos, Thales and Aristeia programmes co-financed by EU-ESF and the Greek NSRF, Greece;
BSF-NSF and MINERVA, Israel;
Norwegian Financial Mechanism 2014-2021, Norway;
NCN and NAWA, Poland;
La Caixa Banking Foundation, CERCA Programme Generalitat de Catalunya and PROMETEO and GenT Programmes Generalitat Valenciana, Spain;
G\"{o}ran Gustafssons Stiftelse, Sweden;
The Royal Society and Leverhulme Trust, United Kingdom.
 
The crucial computing support from all WLCG partners is acknowledged gratefully, in particular from CERN, the ATLAS Tier-1 facilities at TRIUMF (Canada), NDGF (Denmark, Norway, Sweden), CC-IN2P3 (France), KIT/GridKA (Germany), INFN-CNAF (Italy), NL-T1 (Netherlands), PIC (Spain), ASGC (Taiwan), RAL (UK) and BNL (USA), the Tier-2 facilities worldwide and large non-WLCG resource providers. Major contributors of computing resources are listed in Ref.~\cite{ATL-SOFT-PUB-2023-001}.


\clearpage
\printbibliography
 
\clearpage
 
\begin{flushleft}
\hypersetup{urlcolor=black}
{\Large The ATLAS Collaboration}

\bigskip

\AtlasOrcid[0000-0002-6665-4934]{G.~Aad}$^\textrm{\scriptsize 101}$,
\AtlasOrcid[0000-0002-5888-2734]{B.~Abbott}$^\textrm{\scriptsize 119}$,
\AtlasOrcid[0000-0002-7248-3203]{D.C.~Abbott}$^\textrm{\scriptsize 102}$,
\AtlasOrcid[0000-0002-1002-1652]{K.~Abeling}$^\textrm{\scriptsize 55}$,
\AtlasOrcid[0000-0002-8496-9294]{S.H.~Abidi}$^\textrm{\scriptsize 29}$,
\AtlasOrcid[0000-0002-9987-2292]{A.~Aboulhorma}$^\textrm{\scriptsize 35e}$,
\AtlasOrcid[0000-0001-5329-6640]{H.~Abramowicz}$^\textrm{\scriptsize 150}$,
\AtlasOrcid[0000-0002-1599-2896]{H.~Abreu}$^\textrm{\scriptsize 149}$,
\AtlasOrcid[0000-0003-0403-3697]{Y.~Abulaiti}$^\textrm{\scriptsize 116}$,
\AtlasOrcid[0000-0003-0762-7204]{A.C.~Abusleme~Hoffman}$^\textrm{\scriptsize 136a}$
% ...
\bigskip
\\

% ...
$^{29}$Physics Department, Brookhaven National Laboratory, Upton NY; United States of America.\\
% ...
$^{35}$$^{(a)}$Facult\'e des Sciences Ain Chock, R\'eseau Universitaire de Physique des Hautes Energies - Universit\'e Hassan II, Casablanca;$^{(b)}$Facult\'{e} des Sciences, Universit\'{e} Ibn-Tofail, K\'{e}nitra;$^{(c)}$Facult\'e des Sciences Semlalia, Universit\'e Cadi Ayyad, LPHEA-Marrakech;$^{(d)}$LPMR, Facult\'e des Sciences, Universit\'e Mohamed Premier, Oujda;$^{(e)}$Facult\'e des sciences, Universit\'e Mohammed V, Rabat;$^{(f)}$Institute of Applied Physics, Mohammed VI Polytechnic University, Ben Guerir; Morocco.\\
% ...
$^{55}$II. Physikalisches Institut, Georg-August-Universit\"{a}t G\"ottingen, G\"ottingen; Germany.\\
% ...
$^{101}$CPPM, Aix-Marseille Universit\'e, CNRS/IN2P3, Marseille; France.\\
$^{102}$Department of Physics, University of Massachusetts, Amherst MA; United States of America.\\
% ...
$^{116}$Department of Physics, New York University, New York NY; United States of America.\\
% ...
$^{119}$Homer L. Dodge Department of Physics and Astronomy, University of Oklahoma, Norman OK; United States of America.\\
% ...
$^{136}$$^{(a)}$Departamento de F\'isica, Pontificia Universidad Cat\'olica de Chile, Santiago;$^{(b)}$Millennium Institute for Subatomic physics at high energy frontier (SAPHIR), Santiago;$^{(c)}$Instituto de Investigaci\'on Multidisciplinario en Ciencia y Tecnolog\'ia, y Departamento de F\'isica, Universidad de La Serena;$^{(d)}$Universidad Andres Bello, Department of Physics, Santiago;$^{(e)}$Instituto de Alta Investigaci\'on, Universidad de Tarapac\'a, Arica;$^{(f)}$Departamento de F\'isica, Universidad T\'ecnica Federico Santa Mar\'ia, Valpara\'iso; Chile.\\
% ...
$^{149}$Department of Physics, Technion, Israel Institute of Technology, Haifa; Israel.\\
$^{150}$Raymond and Beverly Sackler School of Physics and Astronomy, Tel Aviv University, Tel Aviv; Israel.\\
% ...
$^{a}$ Also Affiliated with an institute covered by a cooperation agreement with CERN.\\
$^{b}$ Also at An-Najah National University, Nablus; Palestine.\\
$^{c}$ Also at Borough of Manhattan Community College, City University of New York, New York NY; United States of America.\\
$^{d}$ Also at Bruno Kessler Foundation, Trento; Italy.\\
$^{e}$ Also at Center for High Energy Physics, Peking University; China.\\
$^{f}$ Also at Centro Studi e Ricerche Enrico Fermi; Italy.\\
% ...

\end{flushleft}


\end{document}